\PassOptionsToPackage{dvipsnames}{xcolor}
\documentclass[aps,prb,reprint,groupedaddress]{revtex4-2}

\usepackage{mathtools, nccmath}
\usepackage{amssymb}
\usepackage{dsfont}
\usepackage[bb=boondox]{mathalfa}
\usepackage{afterpage}
\usepackage[T1]{fontenc}
\usepackage{booktabs}
\usepackage{float}
\usepackage{pgfplots}
\usepackage{tikz}
\usetikzlibrary{decorations.pathmorphing}
\usetikzlibrary{arrows, backgrounds}
\usepackage[caption=false]{subfig}
\usepackage{placeins}
\usepackage{graphicx}
\graphicspath{ {fig-paper-2/} }
\definecolor{almond}{rgb}{0.94, 0.87, 0.8}

\newcommand{\be}{\begin{equation}}
\newcommand{\ee}{\end{equation}}
\newcommand{\bea}{\begin{eqnarray}}
\newcommand{\eea}{\end{eqnarray}}
\newcommand{\Eq}[1]{Eq.\,(\ref{#1})}
\newcommand{\Eqs}[2]{Eqs.\,(\ref{#1}) and (\ref{#2})}
\newcommand{\Eqsss}[3]{Eqs.\,(\ref{#1}), (\ref{#2}), and (\ref{#3})}

\newcommand{\Fig}[1]{Fig.\,\ref{#1}}
\newcommand{\Figs}[2]{Figs.\,\ref{#1} and \ref{#2}}

\newcommand{\Sec}[1]{Sec.\,\ref{#1}}
\newcommand{\App}[1]{Appendix\,\ref{#1}}

\newcommand{\q}{{\bf q}}

\makeatletter
\def\p@subsubsection{}
\makeatother

\newcommand{\tns}[7]{
\draw[style=#7] (-#1*3/4+#6,#2) -- (-#1*3/4+#6,#4);
\draw[style=#7] (-#1*1/4+#6,#3) -- (-#1*1/4+#6,#4);
\draw[style=#7] (#1*1/4+#6,#2) -- (#1*1/4+#6,#5);
\draw[style=#7] (#1*3/4+#6,#3) -- (#1*3/4+#6,#5);
}

\begin{document}

\title{Impact of the phonon environment on the nonlinear quantum-dot cavity QED. \\ II. Analytical approach
}
\author{L.S. Sirkina}
\affiliation{School of Physics and Astronomy, Cardiff University, The Parade, Cardiff CF24 3AA, United Kingdom}
\author{E.A. Muljarov}
\affiliation{School of Physics and Astronomy, Cardiff University, The Parade, Cardiff CF24 3AA, United Kingdom}


\begin{abstract}
The effect of phonons on a nonlinear optical response of a quantum dot-cavity system in quantum strong coupling regime can be accounted for by a fully analytical treatment, provided that the exciton-phonon dynamics is much faster than the exciton-cavity dynamics. Modern experiments involving semiconductor quantum dots embedded in optical microcavities typically meet this criterion.
We find that, for a relatively small exciton-cavity coupling, the effect of phonons is concentrated mainly in the polaron shift of the exciton frequency and reduction of exciton-cavity coupling by the Huang-Rhys factor. We have generalized this result to an arbitrary optical nonlinearity and demonstrated a good agreement with the exact solution in a wide range of temperatures. This generalization provides access to higher rungs of the Jaynes-Cummings ladder, where exact numerical approaches are impractical or even impossible. At larger coupling strengths and low temperatures, our approximation is also in good agreement with the exact solution, which makes it a very useful tool for addressing the phonon contribution to the coherent dynamics of a nonlinear optical system. We demonstrate our results for third-order optical polarization with varying observation time and delay time between excitation pulses in the form of two-dimensional spectra. These spectra provide useful information about the coherent coupling between the exciton and the cavity modes. The presented analytical approach is also compared with another useful approximation, having the form of a matrix product, which is a special case of the asymptotically exact solution in the limit of a short phonon memory time.
\end{abstract}

\maketitle

\section{Introduction}

The Jaynes-Cummings (JC) model~\cite{jaynes1963comparison} has played a very important role in the study of the exciton-photon dynamics in a microcavity-quantum dot (QD) system. Capturing the anharmonic structure of the JC ladder of states can be used as a test for exciton-cavity quantum strong coupling regime~\cite{reithmaier2004strong,yoshie2004vacuum}. The anharmonicity results in important physical applications, including single photon emitters~\cite{pelton2002efficient,gazzano2013bright,press2007photon} and optical switches~\cite{faraon2008coherent,volz2012ultrafast}.

The anharmonicity of the JC ladder can be studied through a nonlinear optical response.
The four-wave mixing (FWM) polarization can be experimentally measured using heterodyne spectral interferometry technique~\cite{langbein2006heterodyne}.
The FWM response was extensively studied in QD(s) inside optical microcavities~\cite{kasprzak2010up,kasprzak2013coherence,albert2013microcavity}.
The role of phonons in the FWM response of QDs was rigorously explored in the absence of a cavity~\cite{muljarov2006nonlinear}. For a QD-cavity system there is an existing perturbative approach~\cite{groll2020four}, where approximate phonon contribution was studied numerically with varying excitation powers.

We explore the effect of phonons on the FWM response analytically and numerically with varying delay times. Two-dimensional (2D) FWM optical spectroscopy~\cite{kasprzak2011coherent} offers useful information about coherent coupling between the exciton and the cavity modes, demonstrating in particular how the coherence is affected by the coupling of the system to the phonon bath. Here we constrain ourselves to a regime where the exciton-cavity dynamics occurs on a much slower timescale than the exciton-phonon dynamics. In this regime, non-Markovian features, which are typically introduced by phonons, are rather minor, and the dynamics can be described in simpler way. At the same time, this regime is very relevant to most of the modern experiments with semiconductor QD embedded in optical cavities~\cite{gerard1996quantum,kasprzak2010up}. A more rigorous treatment of arbitrary parameter regimes is given in~\cite{paper2}.

We present a fully analytical solution, which we call the polaron approximation (PA), here generalized to arbitrary phase channels and orders of optical nonlinearity. This approach is not limited to the number of rungs, which become important in regimes of higher excitation powers~\cite{allcock2022quantum}, where exact numerical treatments are not promising due to large basis sizes, which pose computational issues.
We compare the PA to the nearest neighbor (NN) approximation, the simplest version of the general asymptotically exact approach presented in~\cite{paper2}. This is another useful result, taking a form of simple matrix multiplication. This approach is non-Markovian in nature but is dealing with a rather short memory kernel and is capable of accurately reproducing the phonon broad band in the regime of the fast exciton-phonon dynamics.

\section{Model and Methods}
\label{theory}

\subsection{System, excitation and measurement}
\label{System}

The Hamiltonian of the system consists of two parts,
\begin{align}
    H=H_{\rm JC}+H_{\rm IB}\,,
    \label{hamilt}
\end{align}
where $H_{\bf JC}$ is the JC Hamiltonian  describing the exciton-photon dynamics and $H_{\rm IB}$ is the independent boson (IB) model Hamiltonian which takes into account the coupling of QD excitons to longitudinal acoustic phonons via the deformation potentials,
\begin{align}
    H_{\rm JC}&=\Omega_x d^\dagger d+\Omega_c a^{\dagger} a+g(a^{\dagger} d +d^{\dagger} a)\,,\\
    H_{\rm IB}&=H_{\rm ph}+d^\dagger d V\,, \\
    H_{\rm ph}&=\sum_\q \omega_{q} b^\dagger_{\bf q} b_{\bf q}, \qquad V=\sum_{\bf q} \lambda_{\bf q}(b^\dagger_{-\bf q}+b_{\bf q})\,.
\end{align}	
Here, $g$ is the exciton-cavity coupling strength, $d^\dagger$ ($a^\dagger$) is the creation operator for the  exciton (cavity) mode having a real frequency $\Omega_x$ ($\Omega_c$), $\lambda_{\bf q}$ is the exciton-phonon coupling matrix element, and operator $b_{\bf q}^\dagger$ creates a phonon in mode ${\bf q}$ with frequency $\omega_{q}$. Units of $\hbar=1$ are used throughout the paper.

The full system is excited by a sequence of ultrashort laser pulses. The evolution of the composite system between and after the pulses obeys a master equation of Lindblad type,
\be
i\dot{\rho}=\mathcal{L} \rho=[ H_{\rm JC},\rho]+[ H_{\rm IB},\rho]+i\gamma_c \mathcal{D}[a]+i \gamma_x \mathcal{D}[d]\,, \label{master}
\ee
where the Lindblad dissipator $\mathcal{D}[c]$ is defined by
\begin{equation}
\mathcal{D}[c]\rho=2c\rho c^\dagger-c^\dagger c\rho-\rho c^\dagger c\,,
\label{lindblad}
\end{equation}
and $\mathcal{L}=\mathcal{L}_{\rm JC}+\mathcal{L}_{\rm IB}$ is the Liouvilian superoperator for the whole system.

The density matrix (DM) can be expanded as
\begin{align}
\rho(t)=\sum_{\eta \xi} \rho_{\eta \xi}(t) |\eta \rangle \langle \xi|\,,
\label{dens-mat-decomp}
\end{align}
using the basis defined by
\be
|0,n\rangle = \frac{(a^\dagger)^n}{\sqrt{n!}} |0,0\rangle\,, \quad\quad
|1,n\rangle = \frac{(a^\dagger)^n}{\sqrt{n!}} d^\dagger |0,0\rangle \,,
\label{basis}
\ee
where $|0,0\rangle$ is the absolute ground state of the JC system and the photon number $n$ is a non-negative integer.

\begin{table}[H]
 \begin{tabular}{ l | c | c | c | c }
    \hline
$|\eta \rangle \langle \xi| = j$ & $\hspace{0.4cm} \alpha_j\hspace{0.4cm}$ &  $\hspace{0.4cm}\beta_j\hspace{0.4cm}$ & $\hspace{0.2cm}\alpha_j-\beta_j\hspace{0.2cm}$ & $\hspace{0.2cm}\alpha_j^2-\beta_j^2\hspace{0.2cm}$  \\   \hline
 $|1, \sigma-1\rangle \langle 0, 0| $ & 1 & 0 & 1 & 1\\
 $|0, \sigma\rangle \langle 0, 0| $ & 0 & 0 & 0 & 0\\ \hline
 $|1, n_\sigma-1\rangle \langle 1, n-1| $ & 1 & 1 & 0 & 0\\
 $|1, n_\sigma-1\rangle \langle 0, n| $ & 1 & 0 & 1 & 1\\
 $|0, n_\sigma\rangle \langle 1, n-1| $ & 0 & 1 & -1 & -1\\
 $|0, n_\sigma\rangle \langle 0, n| $ & 0 & 0 & 0 & 0\\
    \hline
  \end{tabular}
  \caption{First column shows the basis elements of the DM required to describe transitions between rungs $n_\sigma=n+\sigma$ and $n$ of the JC ladder, with integer $\sigma>0$. Other columns show the components of vectors $\vec{\alpha}$ and $\vec{\beta}$ and their combinations. The non-zero values of $\alpha_j$ ($\beta_j$) indicate the presence of exciton in each ket (bra) state of the basis element in the first column. Note that for $\sigma<0$, the bra and ket components should be swapped (as done in Table~\ref{tab2}), as well as the values of $\alpha_j$ and $\beta_j$. }
     \label{tab1}
\end{table}

The evolution of the system described by \Eq{master} can then be represented as a superposition of quantum transitions between states of the system belonging to different rungs of the JC ladder. Relevant basis elements of the DM are listed in the first row of Table~\ref{tab1}. The first two elements
correspond to the ground state to rung $\sigma0$ transitions, where $\sigma$ is a positive integer. Transitions between rungs $n$ and $n_\sigma$, where $n_\sigma=n+\sigma$ are described by four other elements.

If we consider the JC evolution alone, treating $\rho$ as a vector, the Liouvillian matrix (with phenomenological decay rates $\gamma_x,\gamma_c$ included) has a general structure described by separate blocks:
\begin{align}
\mathcal{L}_{\rm JC}=
\begin{pmatrix}
   {\mathcal{L}_0} & {\mathcal{M}_{01}}  & {\mathbb{0}}  & {\hdots} \\
    {\mathbb{0}} & {\mathcal{L}_1}  & {\mathcal{M}_{12}}  & {\hdots}\\
    {\mathbb{0}} & {\mathbb{0}}  & {\mathcal{L}_2}  & {\hdots}\\
    {\vdots} & {\vdots}  & {\vdots}  & {\ddots}
\end{pmatrix},
\label{L}
\end{align}
where $\mathcal{L}_0$ is a $2\times2$ block, $\mathcal{M}_{01}$ is a $4\times2$ block, $\mathcal{L}_n$ and $\mathcal{M}_{n,n+1}$ for $n\geqslant 1$ are $4\times4$ blocks, and $\mathbb{0}$ denote blocks of zero elements. While the $\mathcal{M}$-blocks only include the dampings, the $\mathcal{L}$-blocks consist of exciton-photon coupling strength, their real frequencies and the dampings. In the absence of phonons, this JC Liouvilian can be diagonalized analytically~\cite{allcock2022quantum}. The explicit form of the blocks in $\mathcal{L}_{\rm JC}$ is provided in Appendix \ref{appendix:Lgeneral-structure}.

\begin{figure}[t]
\begin{tikzpicture}
\def\co{Blue!60!cyan};
\def\rd{Orange};
 \tikzset{swapaxes/.style = {rotate=90,yscale=-1}}
  
\foreach \x in {8,...,15} {
      \draw[fill=gray,fill opacity=0.5,draw=none] (0,-3+0.4*\x) rectangle ++(3,0.2);
      \draw[fill=\co,fill opacity=0.8,draw=none] (0,-2.8+0.4*\x) rectangle ++(3,0.2);}    
\foreach \x in {0,...,8} {
      \draw[fill=\co,fill opacity=0.8,draw=none] (0,-3+0.4*\x) rectangle ++(3,0.2);
      \draw[fill=gray,fill opacity=0.5,draw=none] (0,-2.8+0.4*\x) rectangle ++(3,0.2);}
\draw[fill=lightgray,draw=none] (0,0.0) rectangle ++(3,0.8);
\draw [fill=darkgray] (1.45,0.4) circle [radius=0.1];

\begin{scope}[shift={(-1.4,-3.02)}]
    \begin{axis}[
      xmin=-2,xmax=2,
      ymin=-3,ymax=3,
      grid=none,axis x line=none,axis y line=none,samples = 1000,swapaxes
      ]
      \addplot[yellow,line width=0.8mm] {1.5*exp(-2*(x)^2)*cos(deg(13.5*(x)))};
    \end{axis}
     \end{scope}
\begin{scope}[shift={(0,0.45)}]
\draw[<->, line width=1.2mm, darkgray] (1.45,4.05) -- (1.45,2.95);
\draw[-, line width=1.8mm, darkgray] (1.45,3.9) -- (1.45,3.1);
\draw[<->, line width=1mm, yellow] (1.45,4) -- (1.45,3) ;
\draw node at (1,3.8) {$O$} node at (1,3.2) {$Q$};   
\end{scope}  
\tikzset{tstyle/.style={line width=0.5mm,<-,\co,opacity=0.4},
tfstyle/.style={line width=0.5mm,<->,\co,opacity=0.8}}
\begin{scope}[shift={(4.5,-3)}]
\def\y{0.272};\def\s{0.774};\def\x{1.0}; 
\def\a{0};\def\b{2-\y};\def\c{2+\y};\def\d{4-\y*1.41};\def\e{4+\y*1.41};\def\f{6-\y*1.73};\def\g{6+\y*1.73};
\foreach \v/\l in {\a/ $0$,\b/ $LP_1$,\c/ $UP_1$,\d/ $LP_{N-1}$,\e/ $UP_{N-1}$,\f/ $LP_N$,\g/ $UP_N$}{
\draw[<->] (-\x-0.1,\f)--(-\x-0.1,\g) node at (-\x/1.8,6){$2\tilde{g}\sqrt{N}$};
\draw (-\x,\v) -- (\x*5/4,\v);
\node at (\x*5/4+0.6,\v+0.1) {\l};
}
\draw[line width=0.5mm,<->,\rd] (-\x*2/4,\a) -- (-\x*2/4,\b);
\draw[line width=0.5mm,<->,\rd] (\x*0/4,\a) -- (\x*0/4,\c);
\tns{\x}{\c}{\b}{\c+0.4}{\c+0.4}{0}{tstyle};
\tns{\x}{\d}{\e}{\d-0.5}{\d-0.5}{\x/4}{tstyle};
\tns{\x}{\e}{\d}{\f}{\g}{0}{tfstyle};
\tns{\x}{\g}{\f}{\g+0.4}{\g+0.4}{\x/4}{tstyle};

\draw node at (\x*1/8,\c+0.8) {$\vdots$} node at (\x*5/4+0.6,\c+0.8) {$\vdots$};
\draw node at (\x*1/8,\g+0.8) {$\vdots$} node at (\x*5/4+0.6,\g+0.8) {$\vdots$};
\def\y{0.352}; 
\end{scope}
\end{tikzpicture}

\begin{tikzpicture}
\def\co{Blue!60!cyan};
\def\rd{Orange};
\draw [thick,->,black](0.5,0) -- (8.5,0); 

\draw[line width=0.5mm,->,\rd] (1,0) -- (1,0.5);
\draw[line width=0.5mm,->,\co,opacity=0.5] (3.5,0) -- (3.5,0.5);
\draw[line width=0.5mm,->,\co,opacity=0.8] (3.55,0) -- (3.55,0.5);
\draw[thick,-,black] (7.5,-0.1) -- (7.5,0.1);

\draw node at (1.5,0.25) {$ Q^{(\mathrm{I})} $};
\draw node at (4,0.25) {$ Q^{(\mathrm{II})} $};

\draw node at (0.9,-0.25) {$ -\tau $};
\draw node at (3.5,-0.25) {$ 0 $};
\draw node at (7.5,-0.25) {$ t $};
\end{tikzpicture}
\caption{Top left: A schematic diagram of semiconductor QD embedded in a micropillar. The system is excited by laser pulses and the response is measured through the top facet of the micropillar. Top right: Energy level structure  of the JC system, forming a ladder of states with characteristic splitting $2\tilde{g}\sqrt{N}$ (where $\tilde{g}$ is the renormalized vacuum Rabi splitting) between the lower polariton (LP) and the upper polariton (UP) states of the $N$-th rung.
Only the neighboring rungs are connected by transitions ($|\sigma|=1$) which are excited by the first (orange) and the second (blue) pulses. Bottom: Pulsed excitation scheme used in the ${\cal N}$-wave mixing. }
\label{excitation}
\end{figure}
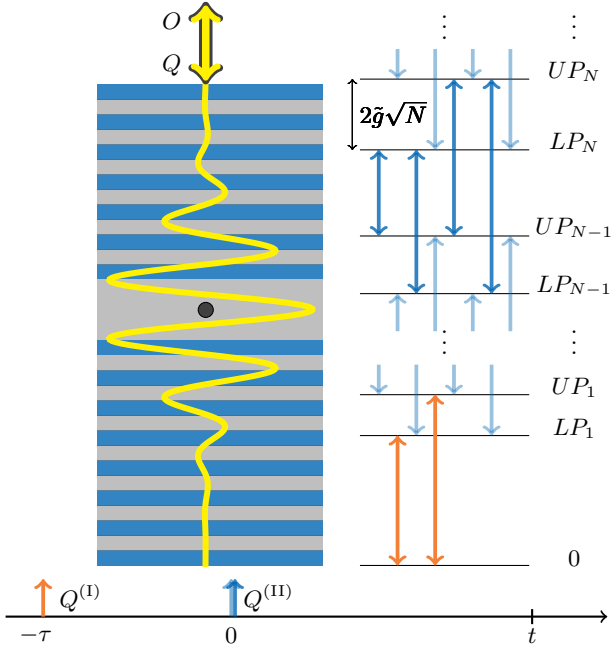

The excitation of the system by a sequence of ultrashort pulses can be expressed by the
interaction operator
\be
 \mathcal{Q}(t)= - \sum_j \delta(t-t_j) \left[ \mathcal{E}_j \mu_{c_j} c_j^\dagger + \mathcal{E}_j^\ast \mu_{c_j}^\ast c_j\right]\,,
     \label{interactExcitationField}
\ee
where $\mu_{c_j}$ is the effective dipole moment and $\mathcal{E}_j$ is the pulse area.
The mode of the JC system to which the external field is coupled, will be referred to as the excitation channel. This can be exciton $c_j^\dagger=d^\dagger$ or cavity $c_j^\dagger=a^\dagger$ or a mixture of both. While the formalism presented below is valid for an arbitrary number of pulses, we focus in this paper on the degenerate ${\cal N}$WM, generated by two pulses of arbitrary strength, separated by the delay time $\tau$ and described by $\vec{Q}^\mathrm{(I)}$ and ${Q}^\mathrm{(II)}$ (their explicit form is given in Appendix \ref{appendix:OQ}). This is schematically shown at the bottom of \Fig{excitation}. We furthermore use the phase selection procedure described in Ref.~\cite{allcock2022quantum}. Namely, pulses I and II introduce, respectively, phases $\Phi=\sigma_\mathrm{I} \Phi_\mathrm{I}$ and $\Phi=\sigma_\mathrm{I} \Phi_\mathrm{I}+\sigma_\mathrm{II} \Phi_\mathrm{II}$ of the selected channel, where $\Phi_j$ is the phase of the complex pulse area $\mu_{c_j}\mathcal{E}_j=|\mu_{c_j}\mathcal{E}_j|e^{i\Phi_j}$. Also, pulses I and II introduce the rung distances $\sigma=\sigma_\mathrm{I}$ and $\sigma=\sigma_\mathrm{I}+\sigma_\mathrm{II}$, determining quantum transitions which contribute to the delay- and real-time dynamics, as described by the JC Liouvilian given above. In the degenerate ${\cal N}$WM, ${\cal N}=|\sigma_\mathrm{I}|+|\sigma_\mathrm{II}|+1$, $\sigma_\mathrm{I}+\sigma_\mathrm{II}=1$, and the lowest order signal after both pulses has a factor
$ie^{i\Phi}|\mu_{c_\mathrm{I}}\mathcal{E}_\mathrm{I}|^{|\sigma_\mathrm{I}|}|\mu_{c_\mathrm{II}}\mathcal{E}_\mathrm{II}|^{|\sigma_\mathrm{II}|}$ (which will be dropped), see Ref.~\cite{allcock2022quantum} for details. Below we consider all possible ${\cal N}$WM channels and arbitrary excitation strengths (i.e. arbitrary pulse areas), but provide numerical illustrations in \Sec{Sec:Results} only for the standard FWM with $\sigma_\mathrm{I}=-1$ and $\sigma_\mathrm{II}=2$, focusing on the lowest-order components of the FWM polaization. For the latter, the excitation operators $\vec{Q}^\mathrm{(I)}$ and ${Q}^\mathrm{(II)}$ are given by \Eqs{Q1}{Q2} in \App{appendix:OQ}.

Following both pulses, the DM of the full system including the exciton-phonon interaction can be formally written as
\begin{align}
 \vec{\rho}(t)=& e^{-i\mathcal{L} t} {Q}^\mathrm{(II)}   e^{-i\mathcal{L} \tau}  \vec{Q}^\mathrm{(I)} \vec{\rho}(-\tau), 
 \label{whole-rho}
\end{align}
where $\vec{\rho}(-\tau)$ is a 1x1 vector, with a single element $\rho_\mathrm{\rm ph}$,
\begin{align}
\rho_\mathrm{\rm ph}&=\frac{\exp(-\frac{H_{\rm ph}}{k_B T})}{\mathrm{tr}_{\rm ph}\{ \exp(-\frac{H_{\rm ph}}{k_B T} )\}}\,.
\label{rho-ph-equilib}
\end{align}
The vector components of the DM $\vec{\rho}(t)$ are the expansion coefficients in \Eq{dens-mat-decomp} represented as a linear array, $\mathcal{L}$ is the corresponding full Liouvillian  matrix operator, including the phonon degrees of freedom, and $\vec{\rho}(-\tau)$ is the initial DM of the optically unexcited system in thermal equilibrium.
The full optical polarization then has the form
\begin{align}
P(t)= \vec{O} \cdot \vec{\rho}(t)\,,
 \label{Pdef}
\end{align}
where $\vec{O}$ is the vector corresponding to the observation channel, which is given explicitly
in \App{appendix:OQ}. We emphasize that $\vec{\rho}(t)$ in \Eqs{whole-rho}{Pdef} is not the full density matrix but only its part which contributes to the response in a selected channel, see Ref.~\cite{allcock2022quantum} for more details.

\subsection{Nearest Neighbor approximation}

Following Ref.~\cite{paper2}, we introduce a discretization of both the delay and observation times,
$\tau=N_\mathrm{I}\Delta \tau$ and $t=N_\mathrm{II}\Delta t$, respectively, where $N_\mathrm{I}$ and $N_\mathrm{II}$ are the numbers of discrete steps. Using Trotter's decomposition, we obtain in the nearest-neighbor (NN) approximation
\be
P(t,\tau)=
\vec{O} \cdot E^{\mathrm{(II)}} \left[{\mathcal{G}^{\mathrm{(II)}}}\right]^{N_\mathrm{II}-1}
\mathcal{G}^{\mathrm{(I-II)}} \left[{\mathcal{G}^{\mathrm{(I)}}}\right]^{N_\mathrm{I}-1} {M}^{\mathrm{(I)}} \vec{Q}^{\mathrm{(I)}}\,,
\label{P-NN-delay}
\ee
where
\begin{align}
\mathcal{G}^{\mathrm{(\zeta)}}_{ij}=& {M}^{\mathrm{(\zeta)}}_{ij} \exp\left({\mathcal{K}^\mathrm{(\zeta)}_{jj}(0)  +2\mathcal{K}^{\mathrm{(\zeta)}}_{ij}(1)}\right),
\label{G1-mat}
\\
\mathcal{K}^{\mathrm{(\zeta)}}_{ij}(l)&=(\alpha_{i}^{\mathrm{(\zeta)}}-\beta_{i}^{\mathrm{(\zeta)}})
\left(\alpha_{j}^{\mathrm{(\zeta)}} R_l^{\mathrm{(\zeta)}}-\beta_{j}^{\mathrm{(\zeta)}} {R_l^{\mathrm{(\zeta)}}}^\ast\right),
\label{cumulant-mat}
\end{align}
with $\zeta$ taking the values I or II,
\begin{align}
\mathcal{G}^{\mathrm{(I-II)}}_{ij}=& [{M}^{\mathrm{(II)}} Q^{\mathrm{(II)}}]_{ij} \exp\left({\mathcal{K}^\mathrm{(I)}_{jj}(0)  +2\mathcal{K}^{\mathrm{(I-II)}}_{ij}(1)}\right),
\label{G1-2-mat}
\\
\mathcal{K}^{\mathrm{(I-II)}}_{ij}(1)&=(\alpha_{i}^{\mathrm{(II)}}-\beta_{i}^{\mathrm{(II)}})
\left(\alpha_{j}^{\mathrm{(I)}} R_1^{\mathrm{(I-II)}}-\beta_{j}^{\mathrm{(I)}} {R_1^{\mathrm{(I-II)}}}^\ast\right),
\label{cumulant-1-2-mat}
\end{align}
$E_{ij}^{\mathrm{(II)}}= \delta_{ij} e^{\mathcal{K}^{\mathrm{(II)}_{jj}(0)}}$, and matrices ${M}$ are defined as
\be
    {M}^{\mathrm{(\zeta)}}=
    e^{-i \mathcal{L}^{\mathrm{(\zeta)}}_{JC} \Delta t_{\mathrm{\zeta}}}\,,
\label{M-matrix}
\ee
with $\Delta t_{\mathrm{I}}=\Delta \tau $ and $\Delta t_{\mathrm{II}}=\Delta t$.
The cumulant elements required for the NN approximation are given by~\cite{paper2}
\begin{align}
R^{\mathrm{(I)}}_0&=K(\Delta \tau)\,, \nonumber \\
R^{\mathrm{(I)}}_1&=K(2\Delta \tau)/2-R^{\mathrm{(I)}}_0, \nonumber \\
R^{\mathrm{(II)}}_0&=K(\Delta t)\,, \nonumber \\
R^{\mathrm{(II)}}_1&=K(2\Delta t)/2-R^{\mathrm{(II)}}_0,
\nonumber \\
2R^{\mathrm{(I-II)}}_1&=K(\Delta \tau +\Delta t)-R^{\mathrm{(I)}}_0-R^{\mathrm{(II)}}_0\,,
\label{NN-cumulant-elements}
\end{align}
and are all expressed in terms of the cumulant function
\be
K(t)=-\dfrac{1}{2} \int_0^{t} dt_1 \int_0^{t} dt_2 {D}(t_1-t_2),
\label{Kt}
\ee
where ${D}(t_1-t_2)=\langle T V(t_1) V(t_2) \rangle$  is the phonon propagator.

Let us finally note that the Liouvillian matrices $\mathcal{L}^{\mathrm{(I)}}_{JC}$ and $\mathcal{L}^{\mathrm{(II)}}_{JC}$ and the corresponding vectors $\vec{\alpha}^{\mathrm{(I)}}$, $\vec{\beta}^{\mathrm{(I)}}$, $\vec{\alpha}^{\mathrm{(II)}}$, and $\vec{\beta}^{\mathrm{(II)}}$ are generally defined by \Eq{L} and Table~\ref{tab1}. The specific form of the matrices $\mathcal{L}^{\mathrm{(I)}}_{JC}$ and $\mathcal{L}^{\mathrm{(II)}}_{JC}$ is determined by the selected channels of the ${\cal N}$WM as discussed in \Sec{System}, which in turn determine the distances $\sigma$ between the rungs of the JC ladder and the corresponding transitions involved in the ${\cal N}$WM coherences: $\sigma=\sigma_\mathrm{I}$ for  $\mathcal{L}^{\mathrm{(I)}}_{JC}$ and $\sigma=\sigma_\mathrm{I}+\sigma_\mathrm{II}$ for $\mathcal{L}^{\mathrm{(II)}}_{JC}$. The sign of $\sigma$, in turn, determines the form of the vectors $\vec{\alpha}$ and $\vec{\beta}$ in each case: They should be taken as given in Table~\ref{tab1} for $\sigma>0$ and swapped for $\sigma<0$.

For the FWM channel (with $\sigma_\mathrm{I}=-1$ and $\sigma_\mathrm{II}=2$), which is our main focus for the rest of the paper, the delay dynamics described by $\mathcal{L}^{\mathrm{(I)}}_{JC}$ corresponds to $\sigma=-1$, so that the bra and ket states in the first column in Table~\ref{tab1} should be swapped  (see Table~\ref{tab2}) and then used with $\sigma=1$ for generating the full matrix $\mathcal{L}^{\mathrm{(I)}}_{JC}$. At the same time, $\vec{\alpha}^{\mathrm{(I)}}=\vec{\beta}$ and
$\vec{\beta}^{\mathrm{(I)}}=\vec{\alpha}$, with $\vec{\alpha}$ and $\vec{\beta}$ given in Table~\ref{tab1}.
Matrix $\mathcal{L}^{\mathrm{(II)}}_{JC}$, describing the real-time evolution, is generated by using the basis elements in the first column in Table~\ref{tab1} with  $\sigma=1$, and is provided in \App{appendix:Lgeneral-structure}, while $\vec{\alpha}^{\mathrm{(I)}}=\vec{\alpha}$ and $\vec{\beta}^{\mathrm{(I)}}=\vec{\beta}$.

\subsection{Polaron approximation}

We use the NN result as a starting point for obtaining a fully analytic expression in the long-time limit. Let us first concentrate on a single time region (for example, region II). Omitting the indices indicating the time region for clarity of presentation, we shall evaluate in the long-time limit the product of matrices $E \mathcal{G}^N$, contributing to \Eq{P-NN-delay}. We consider the observation time $t=N\Delta t$ and even the time step $\Delta t$ large compared to the phonon memory time $\tau_{\rm IB}$. For
$t\gtrsim\tau_{\rm IB}$, the cumulant function \Eq{Kt} can be approximated as
\be
K(t)\approx -i \Omega_p t - S\,,
\label{Kasymp}
\ee
where $\Omega_p$ is the polaron shift and $S$ is the Huang-Rhys factor, see~\cite{paper2} for their explicit form. We obtain from \Eqs{NN-cumulant-elements}{Kasymp} the following approximation for the cumulant elements:
\be
R_0\approx -i \Omega_p \Delta t - S\,, \quad\quad
R_1\approx S/2\,,
\label{KR0}
\ee
and then find, using \Eq{cumulant-mat},
\begin{align}
\mathcal{K}_{jj}(0)&\approx-S(\alpha_j-\beta_j)^2-i \Omega_p \Delta t(\alpha^2_j-\beta^2_j)\,,
\label{K00}
\\
\mathcal{K}_{ij}(1)&\approx S(\alpha_i-\beta_i)(\alpha_j-\beta_j)/2\,.
\label{K01}
\end{align}
According to \Eq{G1-mat}, the matrix $\mathcal{G}$ has the elements
\be
\mathcal{G}_{ij}=M_{ij}e^{\mathcal{K}_{jj}(0)+2\mathcal{K}_{ij}(1)}\,.
\ee
It is convenient to first redefine $\mathcal{G}\to \tilde{\mathcal{G}}$ as
\be
\mathcal{G}=e^{\mathcal{S}/2}\tilde{\mathcal{G}}e^{-\mathcal{S}/2}\,,
\label{Gt}
\ee
by introducing a diagonal matrix $\mathcal{S}$ having the elements
\be
\mathcal{S}_{ij}=\delta_{ij} (\alpha_j-\beta_j)^2 S\,.
\label{S}
\ee
Then, using \Eqs{K00}{K01}, the matrix elements of $\tilde{\mathcal{G}}$ take the form
\be
\tilde{\mathcal{G}}_{ij}\approx M_{ij}e^{-i \Omega_p \Delta t Y_j -S X_{ij}/2}\,,
\label{tGij}
\ee
where
\begin{align}
Y_j&=\alpha^2_j-\beta^2_j\,,
\nonumber\\
X_{ij}&=(\alpha_i-\beta_i-\alpha_j+\beta_j)^2\,.
\label{YX}
\end{align}
In the discretization scheme, the time step $\Delta t$ complies with the Trotter limit and has to be small compared to $\tau_{\rm JC}$, the period of the Rabi rotation in the JC subsystem in the case of zero detuning. Therefore, matrix $M=\exp(-i \mathcal{L}_{\rm JC} \Delta t) $ can be approximated to first order in $\Delta t$ as
\be
M_{ij}\approx \delta_{ij} - i [\mathcal{L}_{\rm JC}]_{ij} \Delta t\,.
\label{M}
\ee
where $[\mathcal{L}_{\rm JC}]_{ij}$ are the matrix elements of $\mathcal{L}_{\rm JC} $.
Expanding also \Eq{tGij} to first order in $\Delta t$, we find, using the fact that $X_{jj}=0$,
\begin{align}
\tilde{\mathcal{G}}_{ij}&\approx e^{-S X_{ij}/2} (1-i\Omega_p \Delta t Y_j ) (\delta_{ij} - i [\mathcal{L}_{\rm JC}]_{ij} \Delta t)
\nonumber\\
&\approx \delta_{ij}- i[\tilde{\mathcal{L}}_{\rm JC}]_{ij} \Delta t
\nonumber\\
&\approx [e^{-i\tilde{\mathcal{L}}_{\rm JC}\Delta t}]_{ij}\,,
\label{Gt2}
\end{align}
where
\be
[\tilde{\mathcal{L}}_{\rm JC}]_{ij}=e^{-S X_{ij}/2} ([\mathcal{L}_{\rm JC}]_{ij}+\Omega_p\delta_{ij}Y_j )
\ee
is a {\it symmetric} matrix. Moreover, it is easy to see from \Eq{YX} and Table~\ref{tab1} that the new matrix $\tilde{\mathcal{L}}_{\rm JC}$ can be obtained from $\mathcal{L}_{\rm JC}$ simply by making a {\it polaron renormalization} of $\tilde{\mathcal{L}}_{\rm JC}$, which consists of the following two elements: (i) renormalization of the coupling constant,
\be
g\to \tilde{g}=g^{-S/2}\,,
\ee
and (ii) renormalization of the QD exciton transition energy,
\be
\Omega_x\to \tilde{\Omega}_x=\Omega_x+\Omega_p\,.
\ee
Finally, noting that $E_{ij}=\delta_{ij}e^{K_{jj}(0)}\approx \delta_{ij}e^{-\mathcal{S}_{jj}}$, see \Eqs{K00}{S}, we obtain
\be
E \mathcal{G}^N = Ee^{\mathcal{S}/2}\tilde{\mathcal{G}}^N e^{-\mathcal{S}/2}
\approx e^{-\mathcal{S}/2}e^{-i\tilde{\mathcal{L}}_{\rm JC} t} e^{-\mathcal{S}/2}\,.
\label{EGN}
\ee

\subsubsection{Zero delay FWM}
Using the above result, \Eq{EGN}, one can immediately obtain a PA for the  FWM polarization at zero delay, which is valid for $t \gtrsim \tau_{\rm IB}$ in the regime $\tau_{\rm JC} \gg\tau_{\rm IB}$ (equivalent to a relatively small coupling strength $g$). For zero delay, the NN approximation \Eq{P-NN-delay} reduces to
\be
P(t,0)=
\vec{O} \cdot E^\mathrm{(II)} \left[{\mathcal{G}^{\mathrm{(II)}}}\right]^{N_\mathrm{II}-1}
{M}^{\mathrm{(II)}} \vec{Q}\,,
\label{P-NN-zero-delay}
\ee
where $\vec{Q}={Q}^\mathrm{(II)} \vec{Q}^\mathrm{(I)}$, see also~\cite{paper2} for details.
We then find the PA for $\tau=0$ and $t \gtrsim \tau_{\rm IB}$:
\begin{align}
P(t,0)=\vec{O} \cdot e^{-\mathcal{S}^\mathrm{(II)}/2} e^{-i \tilde{\mathcal{L}}^\mathrm{(II)}_{\rm JC} t} e^{-\mathcal{S}^\mathrm{(II)}/2} \vec{Q}\,,
\label{FWMLT}
\end{align}
using the fact that $[\mathcal{G}^{-1}M]_{ij}\approx \delta_{ij}$ for $\Delta t\ll \tau_{\rm JC}$, see  \Eqsss{Gt}{M}{Gt2}.

Now, introducing matrices $U$ and $V$ which diagonalize $\tilde{\mathcal{L}}_{\rm JC}$,
\be
\tilde{\mathcal{L}}^\mathrm{(II)}_{\rm JC}=U^\mathrm{(II)}\tilde{\Omega}^\mathrm{(II)}V^\mathrm{(II)}\,,
\ee
where $\tilde{\Omega}^\mathrm{(II)}$ is a diagonal matrix of $\tilde{\omega}^\mathrm{(II)}_j$, the complex eigenvalues of $\tilde{\mathcal{L}}^\mathrm{(II)}_{\rm JC}$. Note that for any form of the Liouvilian for the JC system, its finite or infinite matrix can be diagonalized analytically, as demonstrated in~\cite{allcock2022quantum}. The polarization \Eq{FWMLT} can then be written as
\be
P(t,0)= \theta(t) \sum_j A_j B_je^{-i\tilde{\omega}^\mathrm{(II)}_j t}\,,
\label{Pt0}
\ee
where
\begin{align}
A_j= \sum_k O_k e^{-\mathcal{S}^\mathrm{(II)}_{kk}/2}U^\mathrm{(II)}_{kj}\,,
\quad
B_j= \sum_k V^\mathrm{(II)}_{jk}e^{-\mathcal{S}^\mathrm{(II)}_{kk}/2} Q_{k}\,,
\end{align}
and $\theta(t)$ is the Heaviside function.
Fourier transforming \Eq{Pt0}, we find the FWM spectrum at zero delay:
\be
\bar{P}(\omega)= i\sum_j \frac{ A_j B_j}{\omega-\tilde{\omega}^\mathrm{(II)}_j}\,.
\ee

\subsubsection{Large-delay FWM}

For non-zero delay, the PA can be developed for both the delay and the observation times being sufficiently large, $t,\,\tau \gtrsim \tau_{\rm IB}$. Similarly to \Eq{KR0}, we approximate the relevant cumulant elements as
\begin{align}
R^{\mathrm{(I)}}_0&= -i \Omega_p \Delta\tau - S\,, \nonumber \\
R^{\mathrm{(II)}}_0&= -i \Omega_p \Delta t - S\,, \nonumber \\
R^{\mathrm{(I)}}_1&=R^{\mathrm{(II)}}_1=R^{\mathrm{(I-II)}}_1=S/2\,.
\label{KR}
\end{align}
Then, using \Eq{EGN}, we obtain
\begin{align}
&E^{\mathrm{(II)}} \left[{\mathcal{G}^{\mathrm{(II)}}}\right]^{N_\mathrm{II}-1}
\mathcal{G}^{\mathrm{(I-II)}} \left[{\mathcal{G}^{\mathrm{(I)}}}\right]^{N_\mathrm{I}-1} {M}^{\mathrm{(I)}}
\nonumber\\
\approx & \,\,
e^{-\mathcal{S}^\mathrm{(II)}/2}e^{-i\tilde{\mathcal{L}}^\mathrm{(II)}_{\rm JC} t} e^{-\mathcal{S}^\mathrm{(II)}/2}
\left[\mathcal{G}^\mathrm{(II)}\right]^{-1}\mathcal{G}^\mathrm{(I-II)} \left[E^{\mathrm{(I)}}\right]^{-1}
\nonumber\\
&\times
e^{-\mathcal{S}^\mathrm{(I)}/2}e^{-i\tilde{\mathcal{L}}^\mathrm{(I)}_{\rm JC} \tau} e^{-\mathcal{S}^\mathrm{(I)}/2}
\nonumber\\
= & \,\,
e^{-\mathcal{S}^\mathrm{(II)}/2}e^{-i\tilde{\mathcal{L}}^\mathrm{(II)}_{\rm JC} t}
\tilde{Q}^\mathrm{(II)}
e^{-i\tilde{\mathcal{L}}^\mathrm{(I)}_{\rm JC} \tau} e^{-\mathcal{S}^\mathrm{(I)}/2}\,,
\end{align}
where
\be
\tilde{Q}^\mathrm{(II)}=
e^{-\mathcal{S}^\mathrm{(II)}/2}
\left[\mathcal{G}^\mathrm{(II)}\right]^{-1}\mathcal{G}^\mathrm{(I-II)} \left[E^{\mathrm{(I)}}\right]^{-1}
e^{-\mathcal{S}^\mathrm{(I)}/2}\,.
\ee
To evaluate $\tilde{Q}^\mathrm{(II)}$, we note that
\begin{align}
[e^{-\mathcal{S}^\mathrm{(II)}/2}]_{ij}&=\delta_{ij} e^{-S(\alpha^\mathrm{(II)}_i-\beta^\mathrm{(II)}_i)^2/2}\,,
\nonumber\\
\mathcal{G}^\mathrm{(II)}_{ij}&=[e^{\mathcal{S}^\mathrm{(II)}/2}e^{-i\tilde{\mathcal{L}}^\mathrm{(II)}_{\rm JC} \Delta t}e^{-\mathcal{S}^\mathrm{(II)}/2}]_{ij}\approx \delta_{ij}  \,,
\nonumber\\
M^\mathrm{(II)}_{ij}&=[e^{-i{\mathcal{L}}^\mathrm{(II)}_{\rm JC} \Delta t}]_{ij}\approx \delta_{ij}  \,,
\nonumber\\
\mathcal{G}^\mathrm{(I-II)}_{ij}&\approx
Q^{\mathrm{(II)}}_{ij} e^{-S(\alpha^\mathrm{(I)}_j-\beta^\mathrm{(I)}_j)^2} e^{S(\alpha^\mathrm{(II)}_i-\beta^\mathrm{(II)}_i)(\alpha^\mathrm{(I)}_j-\beta^\mathrm{(I)}_j)}\,,
\nonumber\\
E^{\mathrm{(I)}}_{ij}&=\delta_{ij}e^{K^\mathrm{(I)}_{jj}(0)}\approx \delta_{ij}e^{-S(\alpha^\mathrm{(I)}_j-\beta^\mathrm{(I)}_j)^2}\,,
\nonumber\\
[e^{-\mathcal{S}^\mathrm{(I)}/2}]_{ij}&=\delta_{ij} e^{-S(\alpha^\mathrm{(I)}_j-\beta^\mathrm{(I)}_j)^2/2}\,,
\nonumber
\end{align}
using \Eqsss{G1-2-mat}{cumulant-1-2-mat}{KR} and where necessary the condition $\Delta t,\,\Delta \tau \ll \tau_{\rm JC}$. Then we arrive at
\be
\tilde{Q}^\mathrm{(II)}_{ij}\approx {Q}^\mathrm{(II)}_{ij} e^{-S(\alpha^\mathrm{(II)}_i-\beta^\mathrm{(II)}_i-\alpha^\mathrm{(I)}_j+\beta^\mathrm{(I)}_j)^2/2}\,,
\ee
compare with \Eqs{tGij}{YX}. The PA for the FWM polarization then takes the form
\be
P(t,\tau)= \vec{O} \cdot e^{-\mathcal{S}^{\mathrm{(II)}}/2} e^{-i \tilde{\mathcal{L}}^{\mathrm{(II)}}_{JC} t}  \tilde{Q}^{\mathrm{(II)}}  e^{-i \tilde{\mathcal{L}}^{\mathrm{(I)}}_{JC} \tau} e^{-\mathcal{S}^{\mathrm{(I)}}/2} \vec{Q}^{\mathrm{(I)}}\,,
\label{P-NN-delay2}
\ee
for $t,\tau \gtrsim \tau_{\rm IB}$. Again, diagonalizing the Liouvilian matrices,
\be
\tilde{\mathcal{L}}^\mathrm{(I)}_{\rm JC}=U^\mathrm{(I)}\tilde{\Omega}^\mathrm{(I)}V^\mathrm{(I)}\,,
\quad\quad
\tilde{\mathcal{L}}^\mathrm{(II)}_{\rm JC}=U^\mathrm{(II)}\tilde{\Omega}^\mathrm{(II)}V^\mathrm{(II)}\,,
\ee
\Eq{P-NN-delay2} can be written as a sum of exponentials,
\be
P(t,\tau)= \theta(t)\theta(\tau) \sum_{jl} C_{jl}e^{-i\tilde{\omega}^\mathrm{(II)}_j t}e^{-i\tilde{\omega}^\mathrm{(I)}_l \tau}\,,
\label{Pttau}
\ee
where $\tilde{\omega}^\mathrm{(\zeta)}_j$ are the eigenvalues of $\tilde{\mathcal{L}}^{\mathrm{(\zeta)}}_{JC}$ (and the diagonal elements of the diagonal matrix $\tilde{\Omega}^\mathrm{(\zeta)}$), and
\begin{align}
C_{jl}&=A_j B_l  \sum_{kk'}  V^\mathrm{(II)}_{jk} \tilde{Q}^{\mathrm{(II)}} _{kk'} U^\mathrm{(I)}_{k'l}\,,
\nonumber\\
A_j&= \sum_k O_k e^{-\mathcal{S}^\mathrm{(II)}_{kk}/2}U^\mathrm{(II)}_{kj}\,,
\nonumber\\
B_l&= \sum_k V^\mathrm{(I)}_{lk}e^{-\mathcal{S}^\mathrm{(I)}_{kk}/2} Q^\mathrm{(I)}_{k}\,.
\end{align}
Finally, Fourier transforming \Eq{Pttau} with respect to both the observation and the delay times, we find the 2D FWM spectrum
\be
\tilde{P}(\omega_t,\omega_\tau)= -\sum_{jl} \frac{C_{jl}}{(\omega_t-\tilde{\omega}^\mathrm{(II)}_j)(\omega_\tau-\tilde{\omega}^\mathrm{(I)}_l)}\,.
\label{P2D}
\ee

\section{Results: characterizing the contribution of fast phonon dynamics to the FWM signal}
\label{Sec:Results}

\subsection{Fixed delay $\tau=0$: effect of phonons on time evolution and spectral shape}

We use realistic cavity parameters as in a FWM experiment~\cite{kasprzak2010up} with QDs embedded in a micropillar cavity.
The polaron shift and the Huang-Rhys factor are calculated using 
the exciton-phonon parameters of InGaAs QDs, taken from Refs.~\cite{muljarov2004dephasing,muljarov2005phonon}. These are all given in Tab.\ref{tabparam}.

\begin{table}[H]
 \begin{tabular}{ l | c }
    \hline
exciton-cavity coupling strength & $g=50\,\mu$eV  \\
cavity radiative decay rate & $\gamma_c=30\,\mu$eV  \\
long-time exciton dephasing rate & $\gamma_x=2\,\mu$eV  \\ \hline
deformation potential difference & $D_c-D_v=-6.5$\,eV  \\
QD material density & $\rho_m=5.65$ $\mathrm{cm}^{-3}$\,g  \\
speed of sound in the QD material & $v_s=4.6$\,km $\mathrm{s}^{-1}$  \\
electron / hole confinement radius & $l=3.3$\,nm  \\
    \hline
  \end{tabular}
  \caption{ }
     \label{tabparam}
\end{table}

We focus on a parameter regime, where the two approximate approaches, NN and PA, show good agreement with the exact results. For a direct comparison to the exact solution and an investigation of other parameter regimes see Ref.~\cite{paper2}. Here we present results for the simplest case of a nonlinear response, namely the FWM polarization. We further limit this to low excitation powers, meaning that it is sufficient to consider only the first two rungs of the JC ladder. However, our theory presented in \Sec{theory} is applicable also to higher-order nonlinearities, which were explored in Ref.~\cite{allcock2022quantum} in the absence of phonons. The PA result is very general and is not limited to low excitation powers, which we explore in this section. The PA has an obvious computational advantage due to its analytic form but can only capture the long-time behavior. Thus, it is better suited for a cavity excitation, where the effects of the phonon broad band are minimal, at least for the parameter regime in which this approximation holds. The NN approximation is a numerical approach, which is capable of addressing also the broad band. This approach is the simplest version of the path integral based method~\cite{paper2}, dealing with a minimal memory kernel. It is very easy to implement numerically as it consist of a simple multiplication of matrices.

We first focus on the behavior of the FWM signal in time domain, while keeping the delay fixed.
In order to separate the behavior which occurs on different timescales, a multi-exponential fit has been applied to the long-time part of the numerical data obtained in the NN approach:
\begin{align}
P(t)\approx \sum_j A_j \exp(-i\omega_j t)\,,
\label{fit}
\end{align}
with $A_j$ and $\omega_j$ being complex fit parameters. In accordance with the PA, index $j$ takes two (six) values for the FWM in the case of the exciton (cavity) excitation.
The transition frequencies (see \Fig{excitation}), which have been modified by phonons are represented by the real parts of $\omega_j$ and the corresponding asymptotic decay rates are associated with the imaginary parts of $\omega_j$. The fit \Eq{fit} is represented by circles centered at $\omega_j$ in a complex frequency plane. The magnitudes of the complex amplitudes $A_j$ are proportional to the circle area, while their corresponding phases are given by the color. The fit function \Eq{fit} is Fourier transformed analytically to obtain the FWM spectra, consisting of a coherent superposition of zero-phonon lines (ZPLs) and a broad band. The broad band is obtained by subtracting the fit \Eq{fit} from the NN data and Fourier transforming this difference numerically. The full spectrum in then obtained by adding this broad band part to the analytical Fourier transform of \Eq{fit}.

\begin{figure}[htbp]
\centering
\subfloat[]{\includegraphics[width=0.45\textwidth]{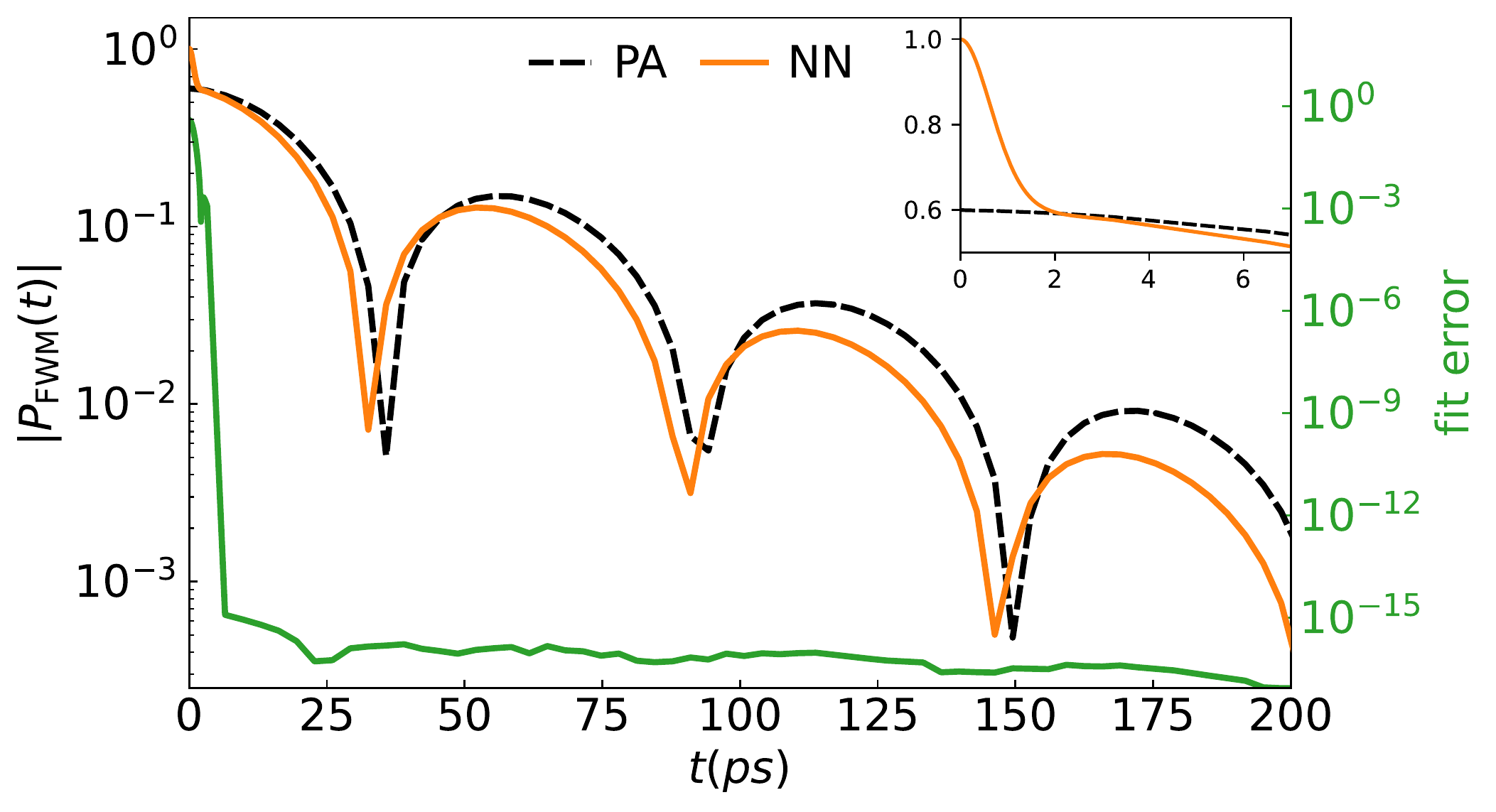}}

\subfloat[]{\includegraphics[width=0.45\textwidth]{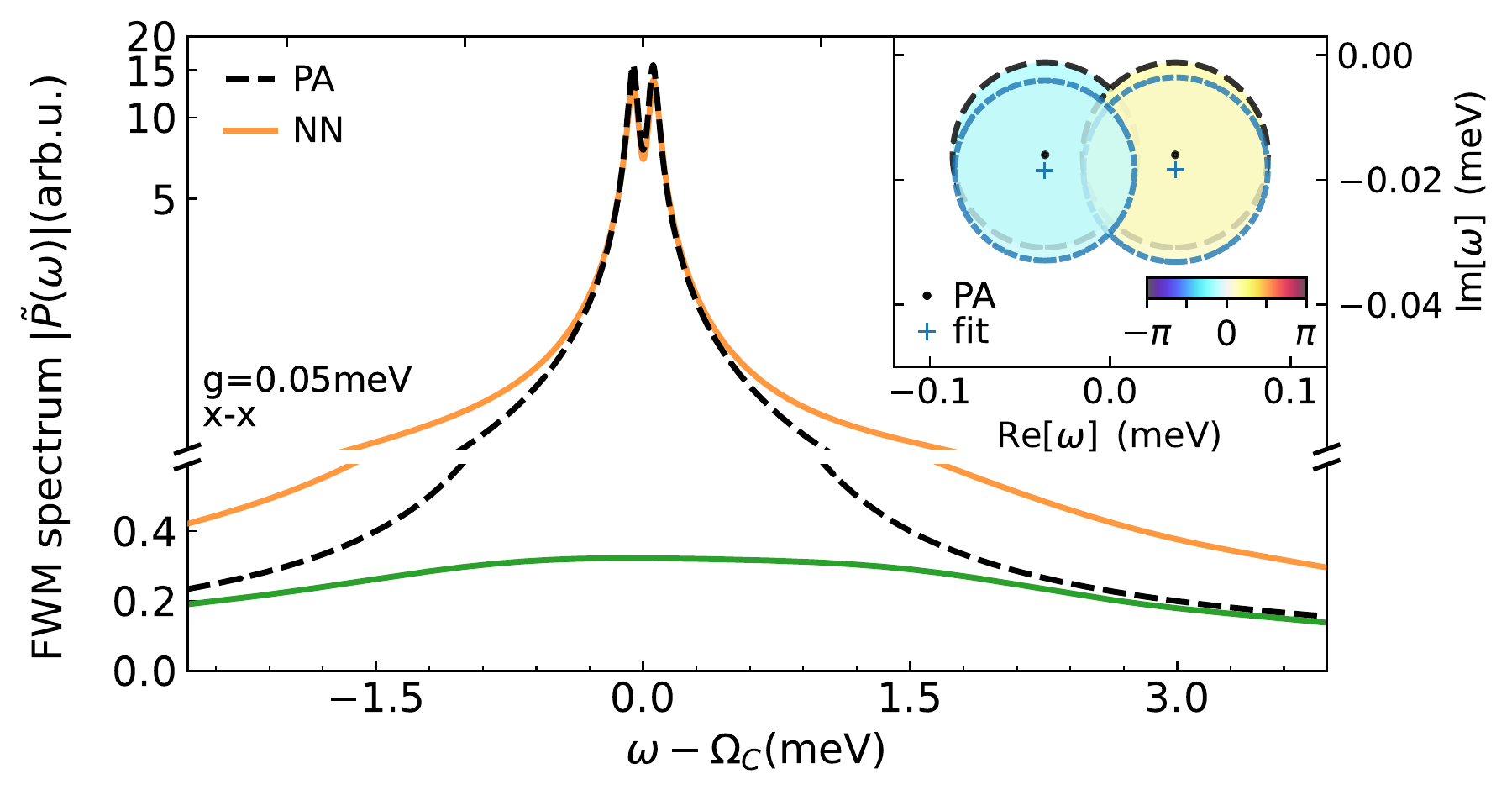}}
\caption{The FWM signal (upper panel) and the corresponding spectrum (lower panel) calculated for $\tau=0$. The excitation and the measurement are done via the exciton mode. The PA (black, dashed) result, the NN result (orange) and the corresponding short-time behavior (green) are shown. In the lower panel, the inset shows the parameters characterizing the long-time multi-exponential behavior. 
The transition frequencies (decay rates) are indicated by the position of circle centers on the horizontal (vertical) axis respectively. The phases are indicated by color. The amplitudes are proportional to circle area. The parameters are $g=0.05$\,meV and $T=50$\,K.}
\label{xx-0-tau}
\end{figure}

\begin{figure}[htbp]
\centering
\subfloat[]{\includegraphics[width=0.45\textwidth]{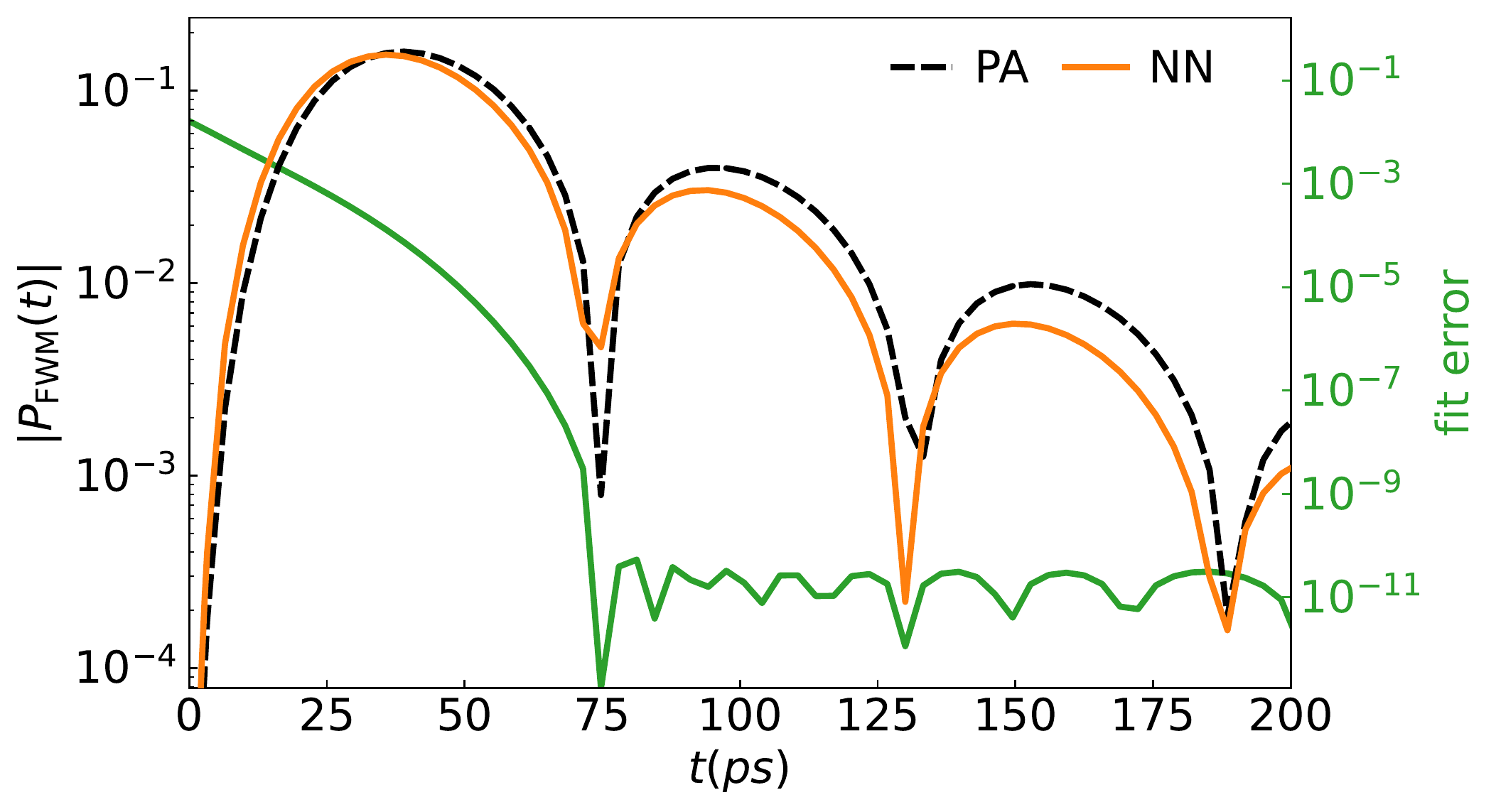}}

\subfloat[]{\includegraphics[width=0.45\textwidth]{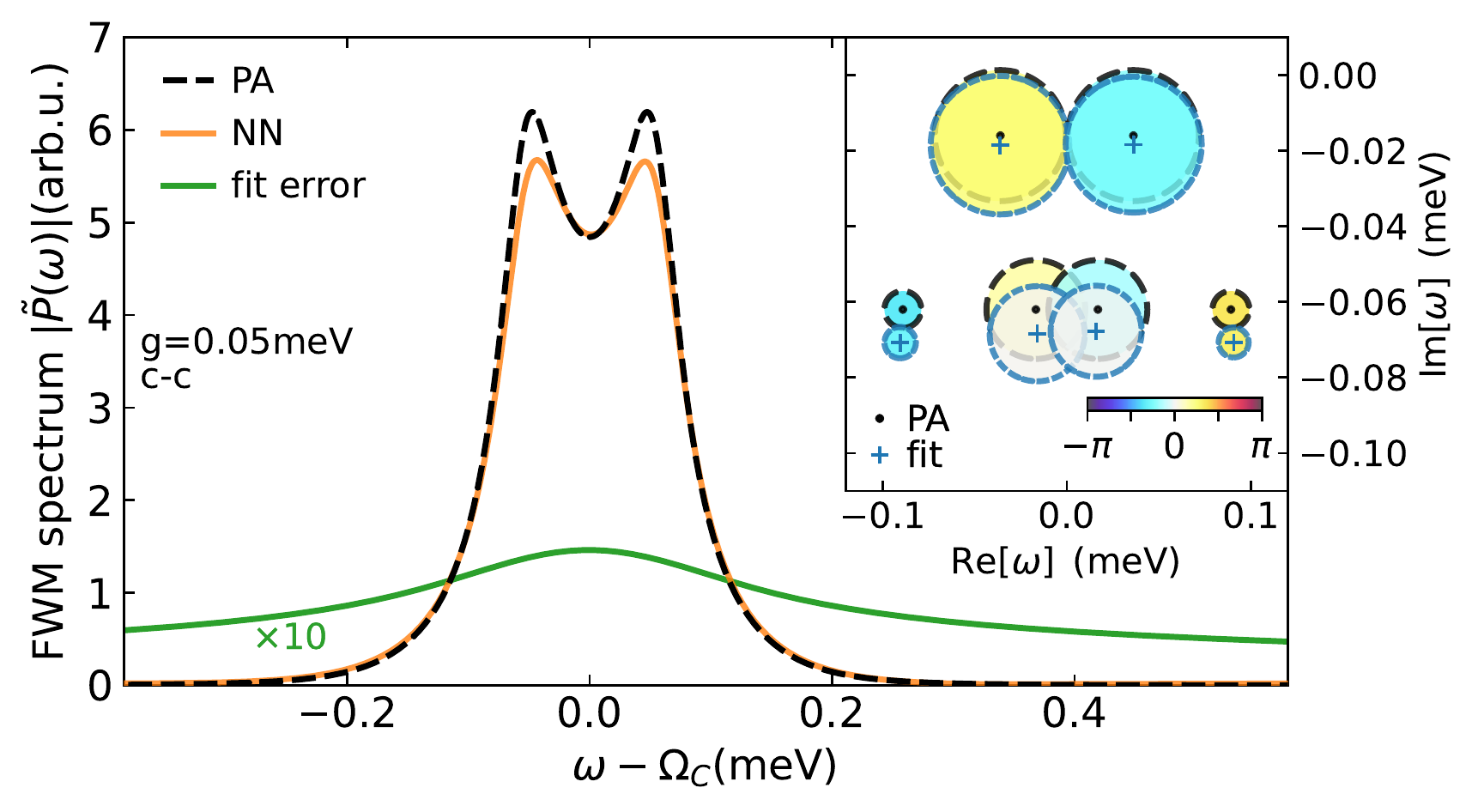}}
\caption{As Fig.\ref{xx-0-tau} but for the excitation and the measurement both done via the cavity mode.}
\label{cc-0-tau}
\end{figure}

Figure \ref{xx-0-tau} shows the FWM polarization and the corresponding spectrum of the system, which is excited and measured via the excitonic mode. The NN result captures the rapid initial decay of the signal, seen over the time period of order $\tau_{\rm IB}$. This results in a broad band covering a wide range of frequencies, much wider than the Rabi splitting, in agreement with the condition $\tau_{\rm IB}\ll \tau_{\rm JC}$. The NN result is close to being exact for $t < \tau_{\rm IB}$ and hence most of the information which is responsible for the broad band can still be captured. In this case of the system excitation via the excitonic mode, the laser pulse creates a phonon cloud around the exciton which is formed on the time scale of $\tau_{\rm IB}$. This results in a significant phonon broad band in the FWM spectrum, which is not captured  by the PA.
If the excitation starts in the cavity mode instead (see \Fig{cc-0-tau}), we do not expect a broad band because the cavity mode does not interact with phonons directly. The height of the broad band is similar for both exciton and cavity channels, but in the latter case the broad band is much narrower which is an evidence of a longer memory time, see~\cite{paper2} for a detailed discussion of this effect.

The difference in the long-time dynamics between the two approaches can be compared directly. This is done in the inset of lower panels of \Figs{xx-0-tau}{cc-0-tau}. For zero delay, the excitonic case is equivalent to the linear polarization. This equivalence can be easily seen in the PA. The main difference between the PA and the NN are the increased decay rates observed in the NN results: the circle centers (blue) are downshifted relative to the PA (black). Another notable feature of the NN result is that the phases for the two inner transitions of the second rung in \Fig{cc-0-tau} are close to zero. At long times within this approximation, the phase information is irreversibly distributed among the continuum of phonon modes. 
This effect is less prominent in the exact solution~\cite{paper2}, where a larger memory kernel allows some of that phase information to return to the exciton-cavity system.

\subsection{Variable delay $\tau \geq 0:$ effect of phonons on coherence in the QD-cavity system}

\begin{figure*}[htbp]
\centering
\subfloat[]{\includegraphics[width=0.25\textwidth]{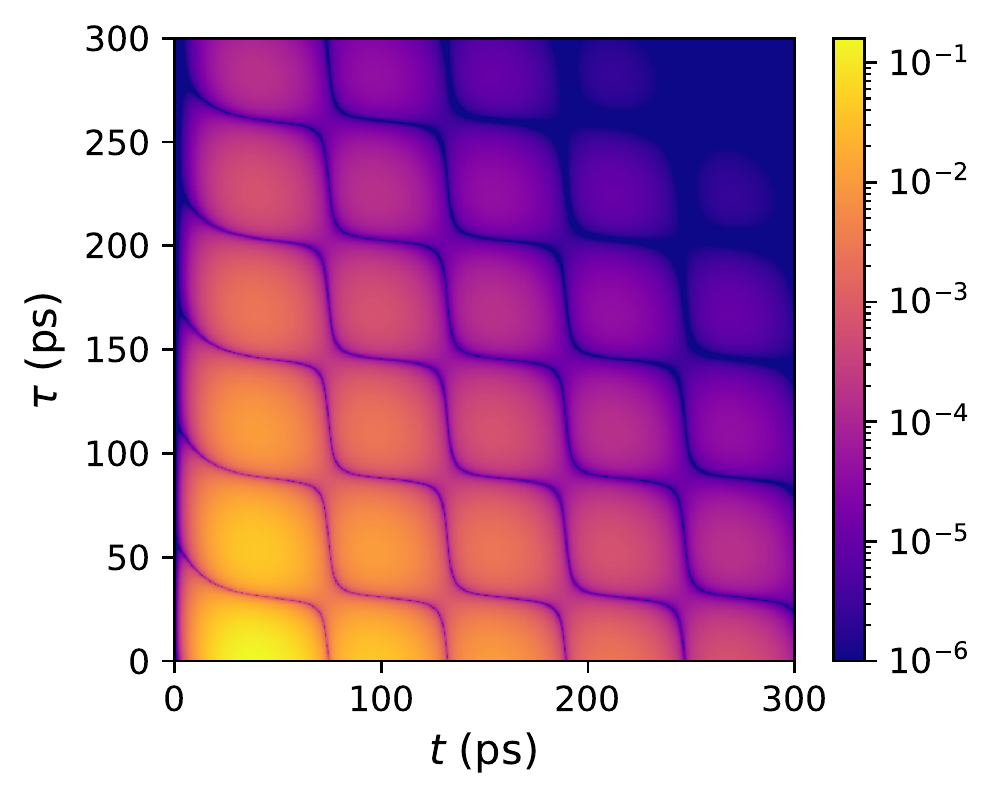}}
\subfloat[]{\includegraphics[width=0.235\textwidth]{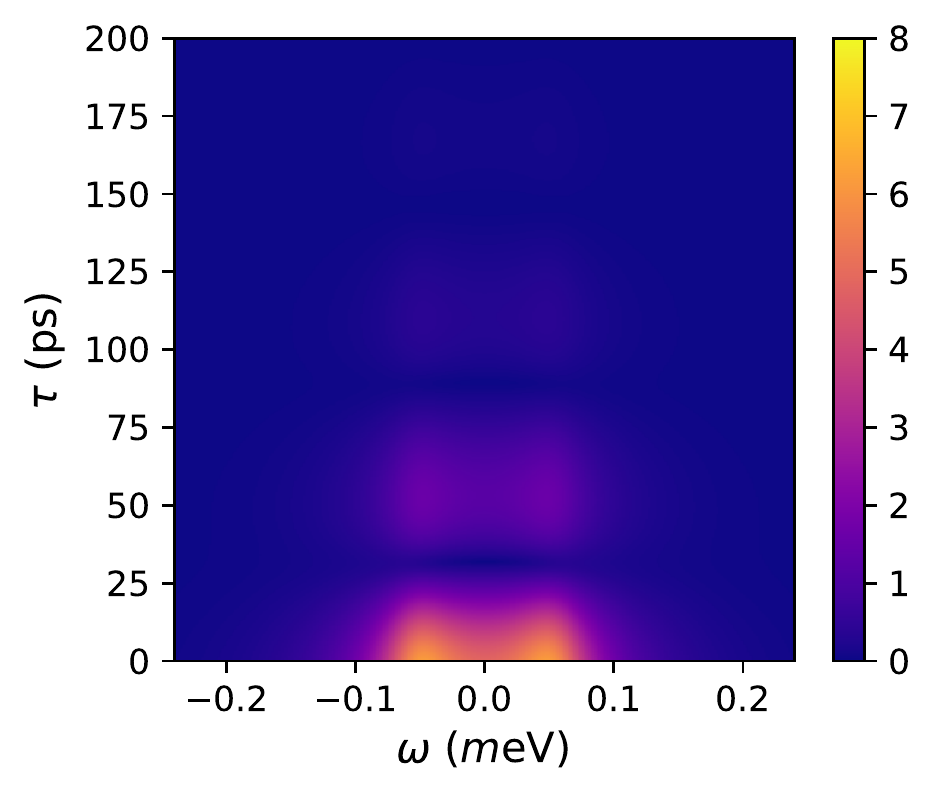}}
\subfloat[]{\includegraphics[width=0.25\textwidth]{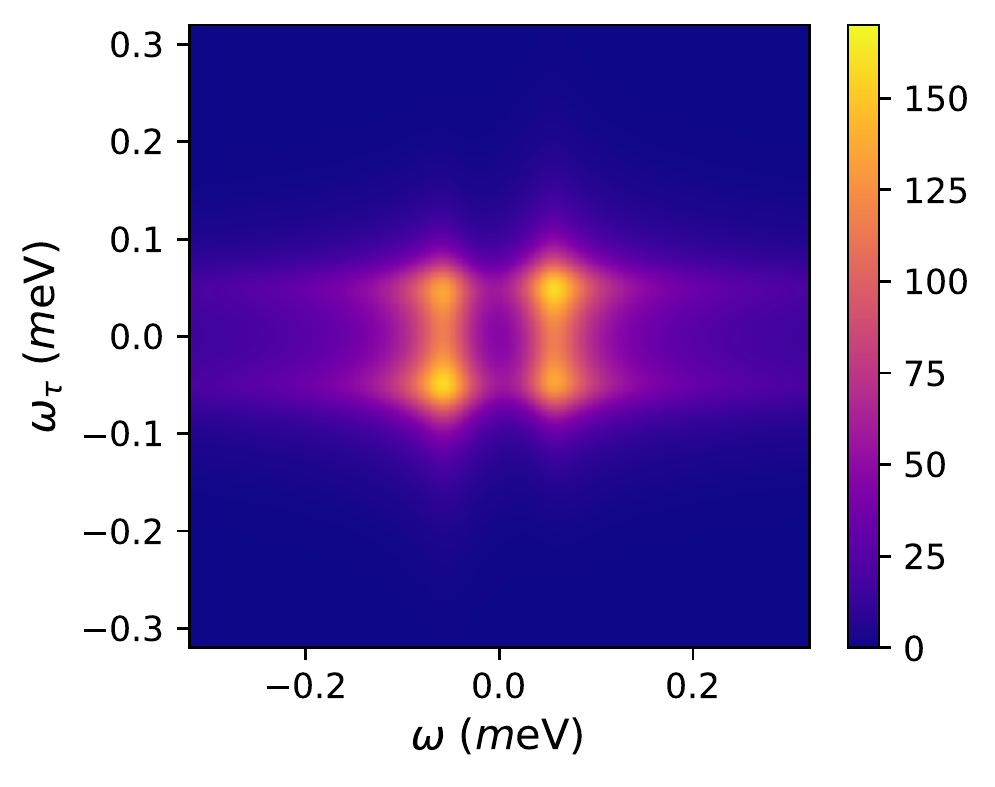}}
\subfloat[]{\includegraphics[width=0.25\textwidth]{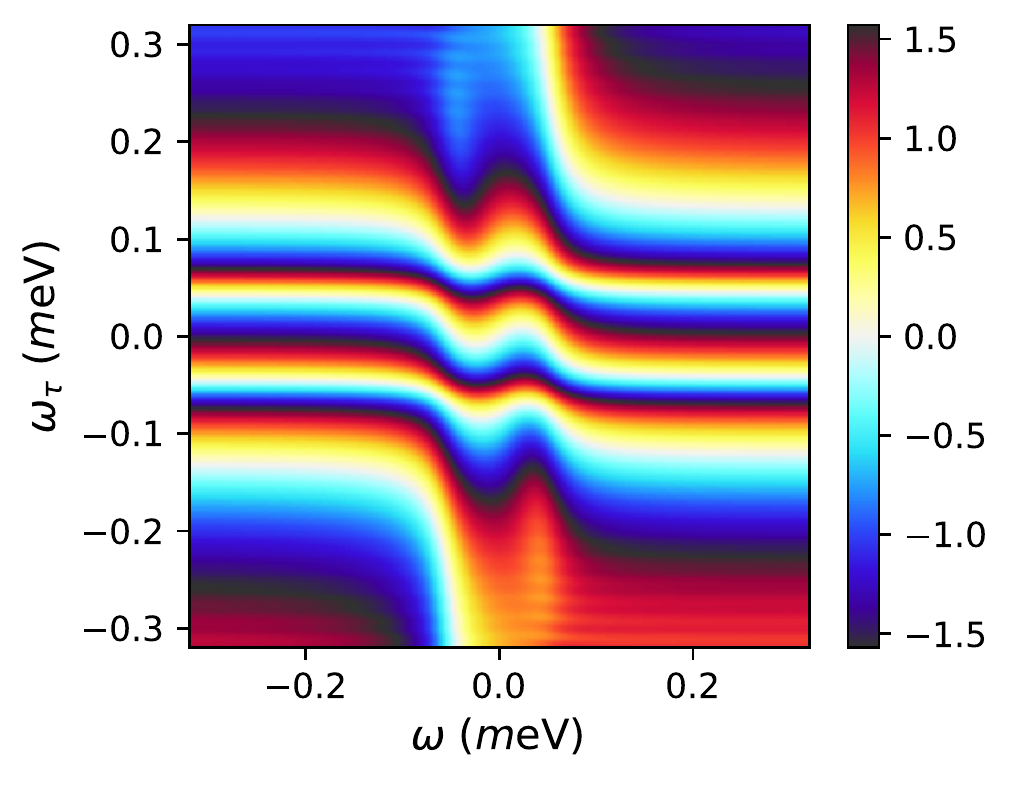}}

\subfloat[]{\includegraphics[width=0.25\textwidth]{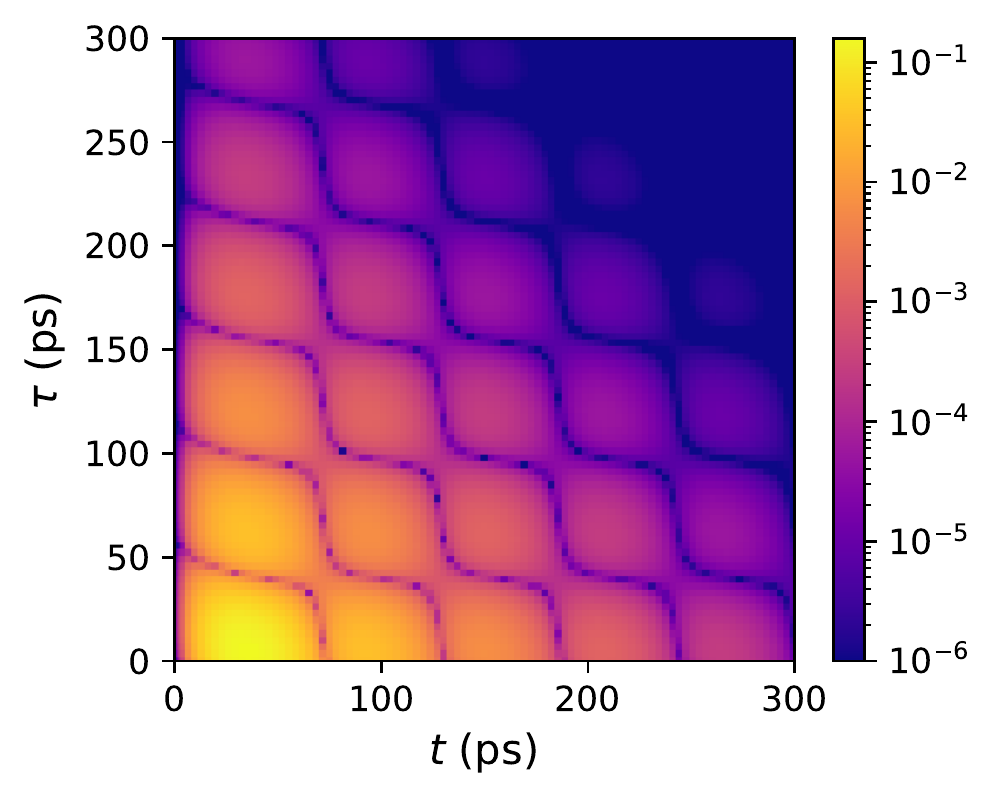}}
\subfloat[]{\includegraphics[width=0.235\textwidth]{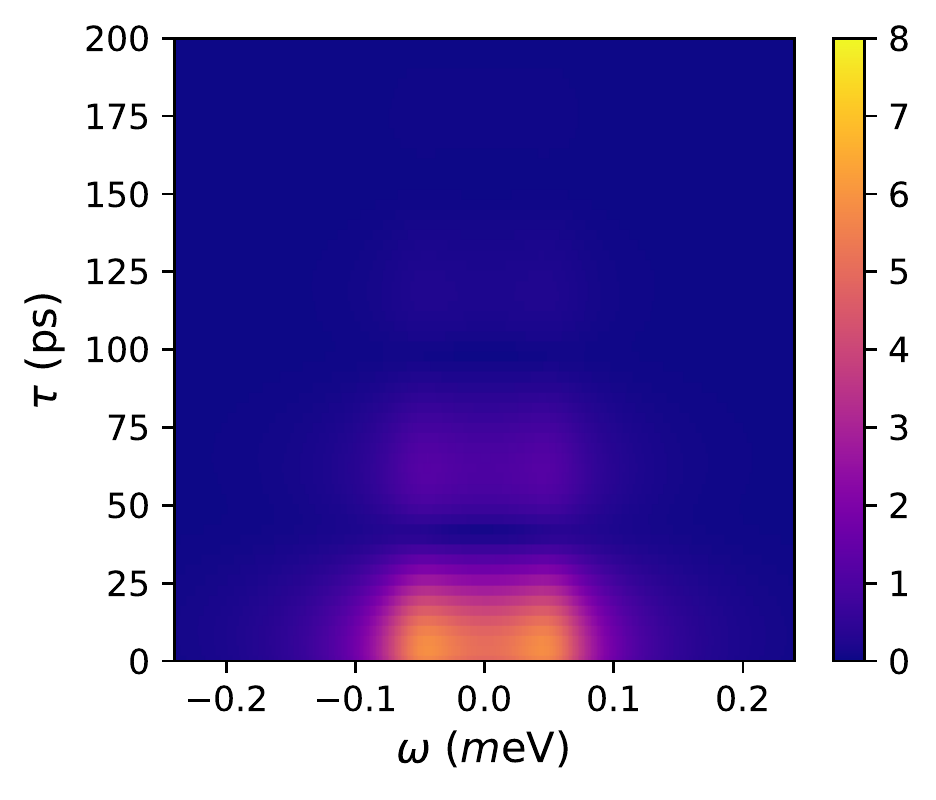}}
\subfloat[]{\includegraphics[width=0.25\textwidth]{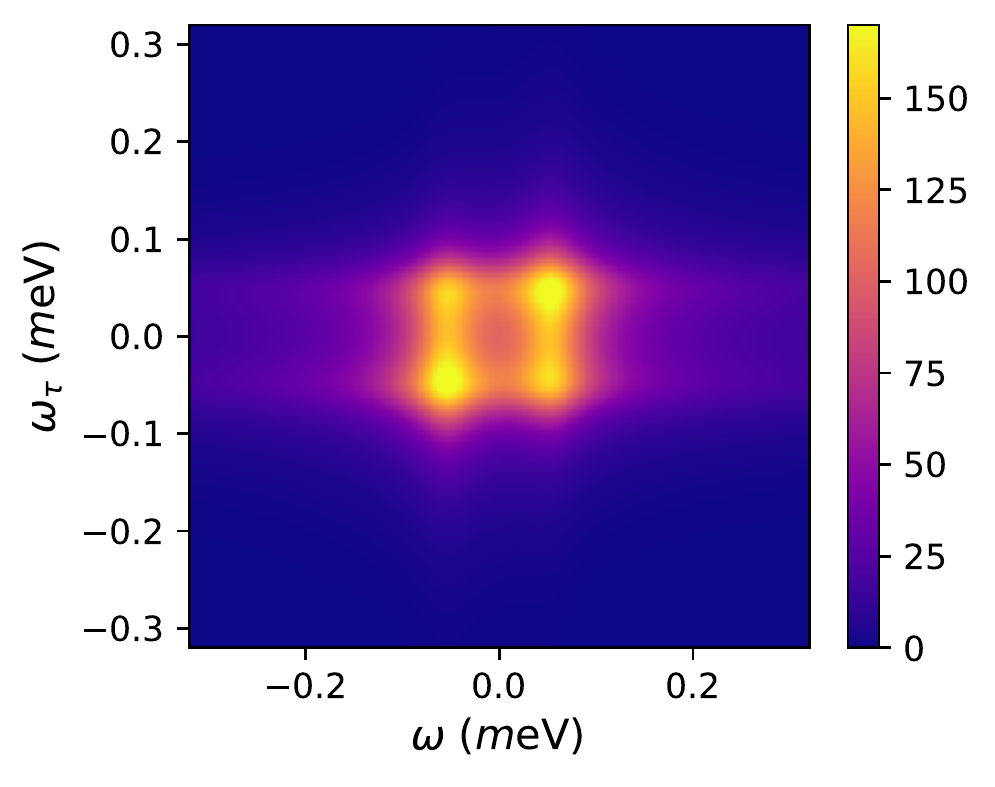}}
\subfloat[]{\includegraphics[width=0.25\textwidth]{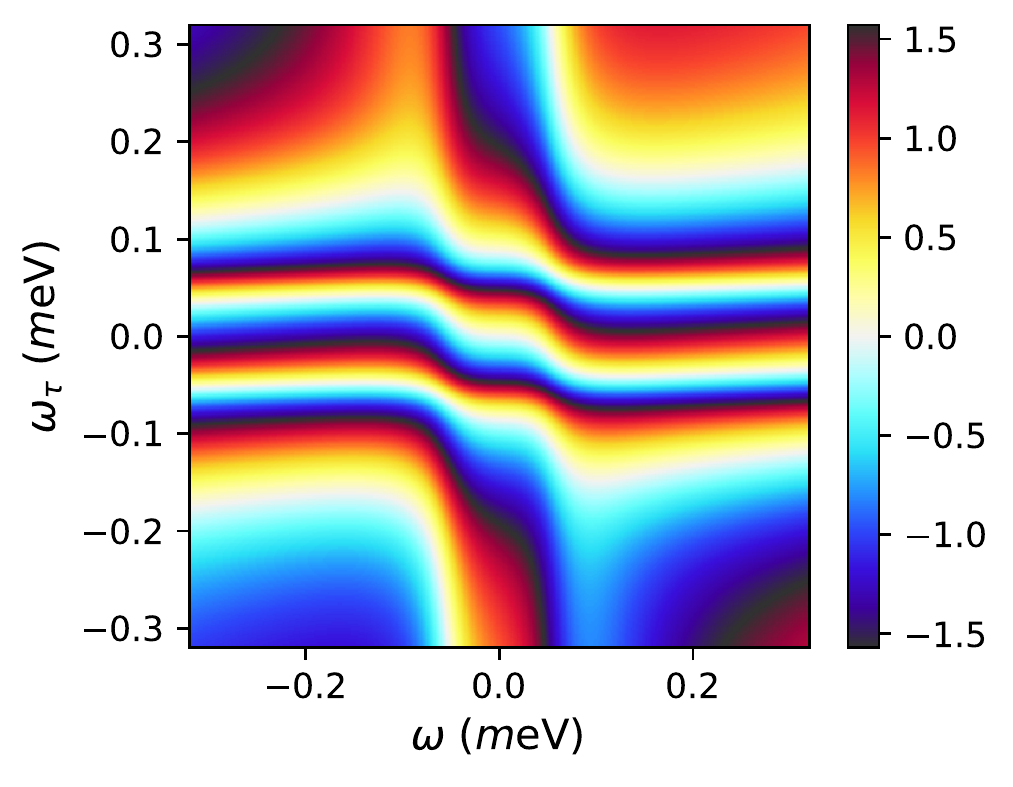}}
\caption{FWM results obtained by the PA (upper panels) and the NN (lower panels) with excitation and measurement via the cavity. The FWM signal as a function of observation time $t$ and delay time $\tau$ with the intensity of the signal shown by color scale (a,e). The Fourier transform is then performed with respect to the observation time (b,f). A 2D Fourier transform with both the delay and the observation times: the absolute values (c,g) and the phases (d,h) are shown separately. The parameters are $g=0.05$\,meV and $T=50$\,K.}
\label{cc-t-tau}
\end{figure*}

For the cavity excitation, corresponding to \Fig{cc-0-tau}, the dependence of the FWM signal on the delay time is shown in \Fig{cc-t-tau} (the excitonic case is shown in Appendix~\ref{appendix:xx-2d-figs}).  The first (second) row of \Fig{cc-t-tau} shows the PA (NN) results.
The first column demonstrates the FWM signal intensity versus the observation and delay times. Horizontally, this figure shows the time evolution of the FWM signal, where the oscillation period is a result of interference of six different transition frequencies of the first two rungs (see \Fig{excitation}).
Fourier transforming it with respect to the observation time, we obtain FWM spectra versus the delay time, which are  presented in the second column of \Fig{cc-t-tau}. For the rather small $g$ used in the calculations, all six transition lines cannot be resolved due to the significant line broadening. In fact, there are only two peaks well seen in the spectra.

As opposed to the excitonic excitation (given in Appendix~\ref{appendix:xx-2d-figs}), where only the first rung coherences are excited, the cavity excitation allows the system to demonstrate a more complicated behavior. 
In the excitonic case only two transitions play a role. By varying the delay between the pulses, it is possible to tune these transitions exactly in antiphase, resulting in no FWM signal.
In the cavity case, the 2D map of the FWM signal in the time domain (Figs.~\ref{cc-t-tau} (a) and (e)) shows that it is not possible to have a complete destructive interference that would produce a vanishing FWM signal at all points in the evolution time. There are four additional transitions, which cannot be brought to an exact antiphase, due to unequal splittings (by $\sqrt{2}$) between the levels of the first and second rung. This is a direct consequence of the nonlinearity. 
The agreement between the PA and NN results in \Fig{cc-t-tau} is very good but not perfect. One of the key differences is a relative shift with delay time (along the vertical axis) between the analytics and the NN results in Figs.~\ref{cc-t-tau} (b) and (f) respectively. This difference is absent when the system is excited via the excitonic channel (see \Fig{xx-t-tau} of Appendix \ref{appendix:xx-2d-figs}). The 2D Fourier transform of the FWM signal in Figs.~\ref{cc-t-tau} (c) and (g) shows four peaks. The diagonal peaks (at $\omega=\omega_\tau$) correspond to the exciton and the cavity states. The other two off-diagonal peaks correspond to coherences within the exciton-cavity system. When the second rung is excited, the 2D spectrum displays a reduced intensity of the off-diagonal peaks as a result of interference between the transitions. The given behavior is not a consequence of the interaction with phonons, it originates from the JC model itself, as can be seen directly from \Eq{P2D}, in which $l=1,\dots, 6$ and $j=1,2$.
The last column in  \Fig{cc-t-tau} shows the phase distribution in the 2D spectra. Following the lines of constant phase one can easily see a difference between the NN and PA, as a result of inclusion of the short-time behavior in the NN (lower panel). Also, the overall intensity of the NN result is greater relative to the PA (see the third column), again, owing to the broad band contribution.

\section{Conclusions}

We have presented a purely analytical polaron approximation (PA) for linear and nonlinear optical response, which takes into account the phonon contribution in the long-time limit, in a regime where the exciton-photon dynamics occurs on a much faster timescale than the exciton-phonon dynamics. We have generalized 
this approximation for an arbitrary channel of optical nonlinearity, focusing on the ${\cal N}$WM and providing numerical illustrations for the more standard FWM polarization. By obtaining the analytical form for the JC Liouvillian modified by the exciton-phonon interaction, we have shown that the effect of phonons manifests itself as a polaron shift of the exciton frequency and a renormalization of the coupling strength by the Huang-Rhys factor. This simple result is proven to hold for any number of rungs, which is a remarkable finding. 
The asymptotic expression for the ${\cal N}$WM polarization has a form analogous to the result previously obtained for the linear polarization, but applies to higher order nonlinearities and arbitrary excitation powers; it has the ability to address all rungs of the JC ladder. While we have illustrated these findings for the FWM with both excitation and measurement done via the same (exciton or cavity) mode, the analytical results are presented for any combination of modes.

We have derived the PA from the nearest-neighbor (NN) approach, which is the simplest form (with the shortest memory time) of the full asymptotically exact solution~\cite{paper2} based on the Trotter decomposition and the cumulant expansion. We have shown in~\cite{paper2} a good agreement of the NN and the PA results with the exact solution.
The NN approach can precisely capture the short-time dynamics and hence model the broad band due to the ability to control the time step size at short times. This feature allows it to excel over the PA in the case of an excitonic excitation. 
For the cavity case the PA is preferred instead, as the fitted NN result appears to exaggerate the broad band and average the phases between some of the first to second rung coherences.
Calculating the optical polarization for arbitrary observation and delay times enabled us to study the coherent coupling using the 2D optical spectroscopy. In the case of an excitonic excitation, the delay time can be tuned to produce a vanishing signal in the time domain as a result of complete destructive interference between the transitions. This is no longer possible for the cavity excitation with low excitation powers. In the 2D FWM spectra we see a reduction in the intensity of the off-diagonal peaks as a result of interference when the coherent coupling is distributed among the two rungs of the JC ladder.

Both the PA and the NN approaches can surpass the computational barrier of numerically exact approaches, dealing with higher rungs of the JC ladder. While the PA results are analogous to making a simple polaron transformation, we have provided a solid theoretical ground that this approximation does not break down for higher order ${\cal N}$WM and can be used to explore the limiting cases of low and high excitation powers. The latter case is associated with the formation of a quantum Mollow quadruplet~\cite{allcock2022quantum}. 
The PA approach presented in this work provides an intuitive understanding and would allow a more thorough analysis of the physical behavior of the exciton-cavity system in the presence of phonons.

\appendix

\section{Explicit structure of the JC Liouvillian}
\label{appendix:Lgeneral-structure}

The explicit form of the different blocks appearing in \Eq{L} and corresponding to transitions (coherences) between rungs $n_\sigma=n+\sigma$ and $n$ is derived by using the basis elements of the DM listed in the first column of Table~\Ref{tab1}. These blocks are given below for a positive integer $\sigma$:
\begin{widetext}
\begin{align}
\mathcal{L}_0=
\begin{pmatrix}
   \omega_x+(\sigma-1)\omega_c & \sqrt{\sigma}g  \\
   \sqrt{\sigma}g & \sigma\omega_c
\end{pmatrix},
\qquad
\mathcal{M}_{01}=
\begin{pmatrix}
   0 & 2i \sqrt{\sigma}\gamma_c  & 0 & 0  \\
   2i \gamma_x & 0  & 0 & 2i\sqrt{\sigma}\gamma_c
\end{pmatrix},
\end{align}

\begin{align}
\small
\mathcal{L}_n=
\begin{pmatrix}
 \omega_x-\omega_x^\ast+(n_\sigma-1)\omega_c-(n-1)\omega_c^\ast & -\sqrt{n}g & \sqrt{n_\sigma}g & 0  \\
 -\sqrt{n}g & \omega_x+(n_\sigma-1)\omega_c-n\omega_c^\ast & 0 & \sqrt{n_\sigma}g \\
   \sqrt{n_\sigma}g & 0 & -\omega_x^\ast+n_\sigma\omega_c-(n-1)\omega_c^\ast & -\sqrt{n}g  \\
   0 & \sqrt{n_\sigma}g & -\sqrt{n}g  & n_\sigma\omega_c-n\omega_c^\ast
\end{pmatrix},
\end{align}
\begin{align}
\mathcal{M}_{n,n+1}
\begin{pmatrix}
  2i \gamma_c\sqrt{n_\sigma n} & 0 & 0 & 0  \\
  0 & 2i \gamma_c\sqrt{n_\sigma (n+1)} & 0 & 0  \\
  0 & 0 & 2i \gamma_c\sqrt{(n_\sigma+1) n} & 0  \\
  2i \gamma_x & 0 & 0 &  2i \gamma_c\sqrt{(n_\sigma+1) (n+1)}
\end{pmatrix},
\end{align}
where for compactness, we have introduced complex frequencies $ \omega_{x,c}= \Omega_{x,c}- i\gamma_{x,c}$, by appending the imaginary decay rates to the real frequencies.

We also provide here explicit form of the JC Liouvillians $\tilde{\mathcal{L}}^{\mathrm{(I)}}_{JC}$ and $\tilde{\mathcal{L}}^{\mathrm{(II)}}_{JC}$, which appear in the PA \Eq{P-NN-delay2} and contribute to the-lowest order FWM polarization (with $\sigma_\mathrm{I}=-1$ and $\sigma_\mathrm{II}=2$):
\begin{align}
\small
\tilde{\mathcal{L}}^{\mathrm{(I)}}_{JC}=
\begin{pmatrix}
   {-\omega_x^\ast-\Omega_p} & {-g e^{-S/2}}\\
   {-g e^{-S/2}} & {-\omega_c^\ast}
\end{pmatrix},
\label{L1-JC}
\end{align}

\begin{align}
\small
\tilde{\mathcal{L}}^\mathrm{(II)}_{\rm JC}=
\left(
\begin{array}{c c|c c c c}
   {\omega_x+\Omega_p} & {g e^{-S/2}} & {0} & {2i \gamma_c} & {0} & {0}  \\
   {g e^{-S/2}} & {\omega_c} & {2 i\gamma_x} & {0} & {0} & {2\sqrt{2} i \gamma_c}  \\
   \midrule
   {0} & {0} & {\omega_c-2 i \gamma_x} & {-g e^{-S/2}} & {\sqrt{2} g e^{-S/2}} & {0}  \\
   {0} & {0} & {-g e^{-S/2}} & {\omega_x- 2 i \gamma_c+\Omega_p} & {0} & {\sqrt{2} g e^{-S/2}}\\
   {0} & {0} & {\sqrt{2} g e^{-S/2}} & {0} & {2\omega_c-\omega_x^*-\Omega_p} & {-g e^{-S/2}} \\
   {0} & {0} & {0} & {\sqrt{2} g e^{-S/2}} & {-g e^{-S/2}} & {\omega_c- 2 i \gamma_c}  \\
\end{array}
\right).
\label{L2-JC}
\end{align}

\end{widetext}
The corresponding matrices ${\mathcal{L}}^{\mathrm{(I)}}_{JC}$ and ${\mathcal{L}}^{\mathrm{(II)}}_{JC}$ can be obtained from \Eqs{L1-JC}{L2-JC} by just setting $\Omega_p=S=0$.

\section{Explicit form of vectors $\vec{O}$ and $\vec{Q}^\mathrm{(I)}$, and matrices $Q^\mathrm{(II)}$, describing excitation and measurement}
\label{appendix:OQ}

Vectors $\vec{O}$ describing the measurement channels are obtained below from the general definition of the optical polarization, $P(t) = \langle {\rho}(t) {O}\rangle= \vec{\rho}(t) \cdot \vec{O}$. For the exciton channel, the operator $O=d$, for which we obtain the expansion
\be
d=\sum_{n=0}^\infty |0, n\rangle \langle 1,n|
\ee
determining the vector $\vec{O}_x$, see Table~\ref{tab2}.
For the cavity channel, $O=a$, and a similar expansion can be written as
\be
a=\sum_{n=0}^\infty \sqrt{n+1} \left( |0, n\rangle \langle 0,n+1| +  |1, n\rangle \langle 1,n+1|\right)
\ee
(see the Supplement in Ref.~\cite{allcock2022quantum},), determining the vector $\vec{O}_c$, which is also shown in Table~\ref{tab2}.

\begin{table}[H]
 \begin{tabular}{ l | c | c | c | c }
    \hline
$|\eta \rangle \langle \xi| = j$ & $\hspace{0.1cm} O_{x,j}\hspace{0.1cm}$ &  $\hspace{0.2cm}O_{c,j}\hspace{0.2cm}$ & $\hspace{0.2cm}Q^\mathrm{(I)}_{x,j}\hspace{0.2cm}$ & $\hspace{0.2cm}Q^\mathrm{(I)}_{c,j}\hspace{0.2cm}$  \\   \hline
 $|0, 0\rangle \langle 1, 0|         $ & 1 & 0 & ${\rm sinc}(2\theta)$ & 0\\
 $|0, 0\rangle \langle 0, 1|         $ & 0 & 1 & 0 & $\exp(-\theta^2)$\\ \hline
 $|1, n-1\rangle \langle 1, n|       $ & 0 & $\sqrt{n}$ & 0 & 0\\
 $|0, n\rangle \langle 1, n|         $ & 1 & 0 & 0 & 0\\
 $|1, n-1\rangle \langle 0, n+1|     $ & 0 & 0 & 0 & 0\\
 $|0, n\rangle \langle 0, n+1| $ & 0 & $\sqrt{n+1}$ & 0 & $\frac{\theta^{2n}\exp(-\theta^2)}{\sqrt{(n+1)!n!}}$\\
    \hline
  \end{tabular}
  \caption{First column shows basis elements of the DM required to describe transitions (coherences) between rungs $n$ and $n+1$ of the JC ladder,  corresponding to $\sigma=-1$. Other columns show elements of the vectors $\vec{O}$ and $\vec{Q}^\mathrm{(I)}$, respectively, for measurement and excitation via the exciton ($x$) or cavity ($c$) mode. Here $\theta=|\mu \mathcal{E}_\mathrm{I}|$ with $\mu=\mu_d$ or $\mu=\mu_a$, respectively, for the exciton or cavity excitation channel, and ${\rm sinc} (x)= \sin x/x$.}
     \label{tab2}
\end{table}

To calculate $\vec{Q}^\mathrm{(I)}_x$, we start with the interaction operator
\be
 \mathcal{Q}_d(t)= -\delta(t) (E^\ast d+E d^\dagger)
     \label{interactExcitationField}
\ee
with $E=\mu_d\mathcal{E}_\mathrm{I}=e^{i\varphi}|E|$, describing excitation of the system via the exciton mode by an ultrashort pulse at $t=0$ with the pulse area $|E|$. Introducing the evolution
operator
\be
U_d=e^{i(E^\ast d+E d^\dagger)}=\cos|E|+i(E^\ast d+E d^\dagger)\frac{\sin|E|}{|E|}\,,
\ee
the instantaneous evolution of the DM due to the pulse is given by
\begin{align}
\rho=&\,U_d \rho_0U_d^\dagger
\nonumber\\
=&\cos^2|E|\rho_0 + \frac{\sin^2|E|}{|E|^2} (E^\ast d+E d^\dagger)\rho_0(E^\ast d+E d^\dagger)
\nonumber\\
&+i\frac{\sin|E|\cos|E|}{|E|} [E^\ast d+E d^\dagger,\rho_0]\,,
\label{Qgenx}
\end{align}
where $\rho_0$ is the DM matrix just before the pulse. Clearly, the only available phase channels past the excitation correspond to $\sigma=0$, 1, and $-1$, as it is clear from the phase factor $e^{i\sigma\varphi}$. Concentrating on the elements of the DM carrying the phase $-\varphi$ and corresponding to $\sigma=-1$, we obtain
\begin{align}
[U_d \rho_0U_d^\dagger]_{-\varphi}=
&-i E^\ast\frac{\sin|E|\cos|E|}{|E|} \rho_0 d
\nonumber\\
=&i^{\sigma} |E|^{|\sigma|}e^{i\sigma\varphi} \frac{\sin(2|E|)}{2|E|} |0, 0\rangle \langle 1, 0| \rho_{\rm ph}\,,
\label{Q1x}
\end{align}
where in the last line we used the fact that the system is fully unexcited before the first pulse, i.e.
\be
\rho_0= |0, 0\rangle \langle 0, 0| \rho_{\rm ph}\,, \quad
\rho_\mathrm{\rm ph}=\frac{\exp(-\frac{H_{\rm ph}}{k_B T})}{\mathrm{tr}_{\rm ph}\{ \exp(-\frac{H_{\rm ph}}{k_B T} )\}}\,.
\label{rho0}
\ee
Leaving aside the common factor $i^{\sigma} |E|^{|\sigma|}e^{i\sigma\varphi}$, we extract  $\vec{Q}^\mathrm{(I)}_x$ from \Eq{Q1x} and list explicitly its elements in Table~\ref{tab2}.

Moving on to matrix ${Q}^\mathrm{(II)}_x$, its explicit form for the standard FWM channel with $\sigma_\mathrm{I}=-1$ and $\sigma_\mathrm{II}=2$  can be also obtained from \Eq{Qgenx}:
\be
[U_d \rho U_d^\dagger]_{2\varphi}= i^2 |E|^2e^{i2\varphi} \left( -\frac{\sin^2|E|}{|E|^2}\right) d^\dagger \rho d\,,
\label{UrUx}
\ee
where now $E=\mu_d\mathcal{E}_\mathrm{II}=e^{i\varphi}|E|$ and $\rho$ is the DM just before the second pulse, which can be written as~\cite{allcock2022quantum}
\begin{align}
\rho= \sum_{n=0}^\infty &\left(\rho^{1,1}_{n,n+1} |1,n\rangle \langle 1, n+1|+
\rho^{0,1}_{n,n} |0,n\rangle \langle 1, n|\right.
\nonumber\\
&\left.+
\rho^{1,0}_{n,n+2} |1,n\rangle \langle 0, n+2|+
\rho^{0,0}_{n,n+1} |0,n\rangle \langle 0, n+1|\right)\,,
\label{rhox}
\end{align}
in accordance with the selected phase channel $\sigma_\mathrm{I}=-1$. Using the fact that
\be
d^\dagger |0,n\rangle \langle 1, n|d^\dagger =|1,n\rangle \langle 0, n|\,,
\ee
we obtain from \Eqs{UrUx}{rhox} an infinite matrix ${Q}^\mathrm{(II)}_x$ which is given in Table~\ref{tab3}. Note that we have again dropped the factor $i^{\sigma_\mathrm{II}} |E|^{|\sigma_\mathrm{II}|}e^{i\sigma_\mathrm{II}\varphi}$, which appears in \Eq{UrUx}.

To calculate $\vec{Q}^\mathrm{(I)}_c$ and ${Q}^\mathrm{(II)}_c$, we use instead the interaction operator
\be
 \mathcal{Q}_a(t)= -\delta(t) (E^\ast a+E a^\dagger)
     \label{interactExcitationFieldC}
\ee
in which $E=\mu_a\mathcal{E}=e^{i\varphi}|E|$ and $\mathcal{E}=\mathcal{E}_\mathrm{I}$ or $\mathcal{E}_\mathrm{II}$, depending on which excitation pulse is being treated. The interaction \Eq{interactExcitationFieldC} generates the evolution operator~\cite{allcock2022quantum}
\be
U_a=e^{i(E^\ast a+E a^\dagger)}=\sum_{nm} e^{i\varphi (n-m)} C_{nm}  |n\rangle \langle m|\,,
\ee
where $|n\rangle$ are the Fock states generated by the cavity operator $a^\dagger$,
\be
C_{nm}=i^{m-n}|E|^{n-m}\sqrt{\frac{m!}{n!}} L_m^{n-m} (|E|^2) e^{-|E|^2/2}=C_{mn}\,,
\ee
and $L_k^n (x)$ are the associated Laguerre polynomials.

Stating from the DM $\rho_0$ of the unexcited system in thermal equilibrium, which is given by \Eq{rho0}, the DM of the system immediately after the excitation pulse has all possible phase channels present. Extracting the part of the DM for the channel corresponding to the phase $\sigma\varphi$, we obtain for $\sigma\geqslant0$
\begin{align}
[U_a \rho_0 U_a^\dagger]_{\sigma\varphi}=& e^{i\sigma\varphi} \sum_{n=0}^\infty C_{n_\sigma0} C_{n0}^\ast |0, n_\sigma\rangle \langle 0, n| \rho_{\rm ph}
\nonumber\\
=&  i^{\sigma} |E|^{|\sigma|}e^{i\sigma\varphi} \rho_{\rm ph}\sum_{n=0}^\infty \frac{|E|^{2n} e^{-|E|^2}}{\sqrt{n!n_\sigma!}} |0, n_\sigma\rangle \langle 0, n|\,,
\label{UrUc-sp}
\end{align}
which is now written in the basis of the JC model. For $\sigma\leqslant0$ we obtain a very similar expression:
\begin{align}
[U_a \rho_0 U_a^\dagger]_{\sigma\varphi}=&
 i^{\sigma} |E|^{|\sigma|}e^{i\sigma\varphi}
  \rho_{\rm ph}
\label{UrUc-sm}\\
& \times \sum_{n=0}^\infty \frac{|E|^{2n} e^{-|E|^2}}{\sqrt{n!(n-\sigma)!}} |0, n\rangle \langle 0, n-\sigma| \,.
\nonumber
\end{align}
For $\sigma=-1$, this gives
\begin{align}
[U_a \rho_0 U_a^\dagger]_{-\varphi}=&
 i^{\sigma} |E|^{|\sigma|}e^{i\sigma\varphi} \rho_{\rm ph}
\label{UrUc-sm1} \\
&
 \times \sum_{n=0}^\infty \frac{|E|^{2n} e^{-|E|^2}}{\sqrt{n!(n+1)!}} |0, n\rangle \langle 0, n+1|\,,
\nonumber
\end{align}
from which we find $\vec{Q}^\mathrm{(I)}_c$ presented in Table~\ref{tab2}.

The matrix ${Q}^\mathrm{(II)}_c$ describing the effect of the second pulse is calculated is a similar way, but starting from the density matrix in the form of \Eq{rhox}. Then, according to \cite{allcock2022quantum},
\be
[U_a \rho U_a^\dagger]_{nn'}^{\nu \nu'}=\sum_{kk'} e^{i\varphi (n-k-n'+k')} C_{nk}C_{n'k'}^\ast \rho_{kk'}^{\nu \nu'}\,,
\ee
where now $E=\mu_a\mathcal{E}_\mathrm{II}=e^{i\varphi}|E|$. Extracting the terms with the phase $2\varphi$, and dropping the common factor $i^2 |E|^2 e^{2i\varphi}$ we find ${Q}^\mathrm{(II)}_c$  for the standard FWM channel (with $\sigma_\mathrm{I}=-1$ and $\sigma_\mathrm{II}=2$), which is presented in Table~\ref{tab4}, where
\be
A_{n,m}=\sqrt{\frac{m!}{n!}} L_m^{n-m} (\theta^2) \theta^{n-m-1} e^{-\theta^2/2}= (-1)^{m-n} A_{m,n}
\label{Anm}
\ee
and $\theta=|E|$. In particular,
\begin{align}
A_{1,0}A_{0,1}=& -A^2_{1,0}= -e^{-\theta^2}\,, \\
A_{N+1,0}A_{N-1,0}=& \frac{\theta^{2N-2}e^{-\theta^2}}{\sqrt{(N+1)!(N-1)!}}\,, \\
A_{N+1,0}A_{N,1}=& \frac{(N-\theta^2)\theta^{2N-2}e^{-\theta^2}}{\sqrt{(N+1)!N!}}\,.
\end{align}
Note also that ${Q}^\mathrm{(II)}_c$ is a symmetric infinite matrix.

\begin{widetext}
\begin{table}[H]
 \begin{tabular}{ l | c  c | c  c c c  }
\ \ \ \ \ \ \  ${Q}^\mathrm{(II)}_x$ &  \ \ $|0, 0\rangle \langle 1, 0|$\ \  &  \ \ $|0, 0\rangle \langle 0, 1|$ \ \ & \ \ $|1, n-1\rangle \langle 1, n|$ \ \ & \ \ $|0, n\rangle \langle 1, n|$\ \  & \ \ $|1, n-1\rangle \langle 0, n+1|$\ \  & \ \  $|0, n\rangle \langle 0, n+1$\ \  \\   \hline
 $|1, 0\rangle \langle 0, 0|$ & $-{\rm sinc}^2(\theta)$ & 0 & 0 & 0& 0& 0\\
 $|0, 1\rangle \langle 0, 0|$ & 0 & 0 & 0 & 0 & 0& 0\\ \hline
 $|1, n\rangle \langle 1, n-1|$  & 0 & 0 & $-{\rm sinc}^2(\theta)$ & 0 & 0& 0\\
 $|1, n\rangle \langle 0, n|$  & 0 & 0 & 0 & 0 & 0& 0\\
 $|0, n+1\rangle \langle 1, n-1|$  & 0 & 0 & 0 & 0 & 0& 0\\
 $|0, n+1\rangle \langle 0, n|$ & 0 & 0 & 0 & 0 & 0& 0\\
    \hline
  \end{tabular}
  \caption{Elements of the infinite matrix ${Q}^\mathrm{(II)}_x$, where $\theta=|\mu_d \mathcal{E}_\mathrm{II}|$ and $n$ is a positive integer labeling rungs of the JC ladder. }
     \label{tab3}
\end{table}
\begin{table}[H]
 \begin{tabular}{ l | c  c | c  c c c  }
\ \ \ \ \ \ \  ${Q}^\mathrm{(II)}_c$ &  \ \ $|0, 0\rangle \langle 1, 0|$\ \  &  \ \ $|0, 0\rangle \langle 0, 1|$ \ \ & \ \ $|1, n-1\rangle \langle 1, n|$ \ \ & \ \ $|0, n\rangle \langle 1, n|$\ \  & \ \ $|1, n-1\rangle \langle 0, n+1|$\ \  & \ \  $|0, n\rangle \langle 0, n+1$\ \  \\   \hline
 $|1, 0\rangle \langle 0, 0|$ & 0 & 0 & 0 & 0& $A_{0,n-1}A_{0,n+1}$& 0\\
 $|0, 1\rangle \langle 0, 0|$ & 0 & $A_{1,0}A_{0,1}$ & 0 & 0 & 0& $A_{1,n}A_{0,n+1}$\\ \hline
 $|1, N\rangle \langle 1, n-1|$  & 0 & 0 & $A_{N,n-1}A_{N-1,n}$ & 0 & 0& 0\\
 $|1, N\rangle \langle 0, n|$  & 0 & 0 & 0 & 0 & $A_{N,n-1}A_{N,n+1}$& 0\\
 $|0, N+1\rangle \langle 1, n-1|$  & $A_{N+1,0}A_{N-1,0}$ & 0 & 0 & $A_{N+1,n}A_{N-1,n}$ & 0& 0\\
 $|0, N+1\rangle \langle 0, n|$ & 0 & $A_{N+1,0}A_{N,1}$ & 0 & 0 & 0& $A_{N+1,n}A_{N,n+1}$\\
    \hline
  \end{tabular}
  \caption{Elements of the infinite matrix ${Q}^\mathrm{(II)}_c$, where $n$ and $N$ are positive integers labeling rungs of the JC ladder, and  $A_{N,n}$ are given by \Eq{Anm} with $\theta=|\mu_a \mathcal{E}_\mathrm{II}|$. }
     \label{tab4}
\end{table}
\end{widetext}

In the lowest-order FWM polarization, all rungs of the JC ladder higher than 2 do not contribute, as the corresponding elements of the excitation and measurement matrices vanish. Then $\vec{O}$, $\vec{Q}^\mathrm{(I)}$, and $Q^\mathrm{(II)}$ reduce to
\begin{align}
\setcounter{MaxMatrixCols}{30}
\vec{O}_x=
\begin{pmatrix}
   {1} \\ {0}  \\   {0}  \\ {1} \\ {0}  \\ {0}
\end{pmatrix},
\qquad
\vec{O}_c=
\begin{pmatrix}
   {0} \\ {1} \\   {1}  \\ {0} \\ {0}  \\ {\sqrt{2}}
\end{pmatrix},
\label{Ovec}
\end{align}
for the measurement, and to
\begin{align}
\vec{Q}^{\mathrm{(I)}}_x=
\begin{pmatrix}
   {1}  \\
	{0}  \\
\end{pmatrix},
\qquad
\vec{Q}^{\mathrm{(I)}}_c=
\begin{pmatrix}
   {0}  \\
	{1}  \\
\end{pmatrix},
\label{Q1}
\end{align}
\begin{align}
{Q}^{\mathrm{(II)}}_x=
\begin{pmatrix}
   {-1} & {0} \\
	{0} & {0} \\
	{0} & {0} \\
   {0} & {0} \\
	{0} & {0} \\
	{0} & {0} \\		
\end{pmatrix},
\qquad
{Q}^{\mathrm{(II)}}_c=
\begin{pmatrix}
   {0} & {0} \\
	{0} & {-1} \\
	{0} & {0} \\
   {0} & {0} \\
	{1/\sqrt{2}} & {0} \\
	{0} & {1/\sqrt{2}} \\	
\end{pmatrix}\,
\label{Q2}
\end{align}
for the excitation. These can all be  obtained directly from Tables \ref{tab2}-\ref{tab4} by setting $\theta=0$.

\section{More results on the FWM: Excitation via the exciton channel}
\label{appendix:xx-2d-figs}

\begin{widetext}
\begin{figure}[H]
\centering
\subfloat[]{\includegraphics[width=0.25\textwidth]{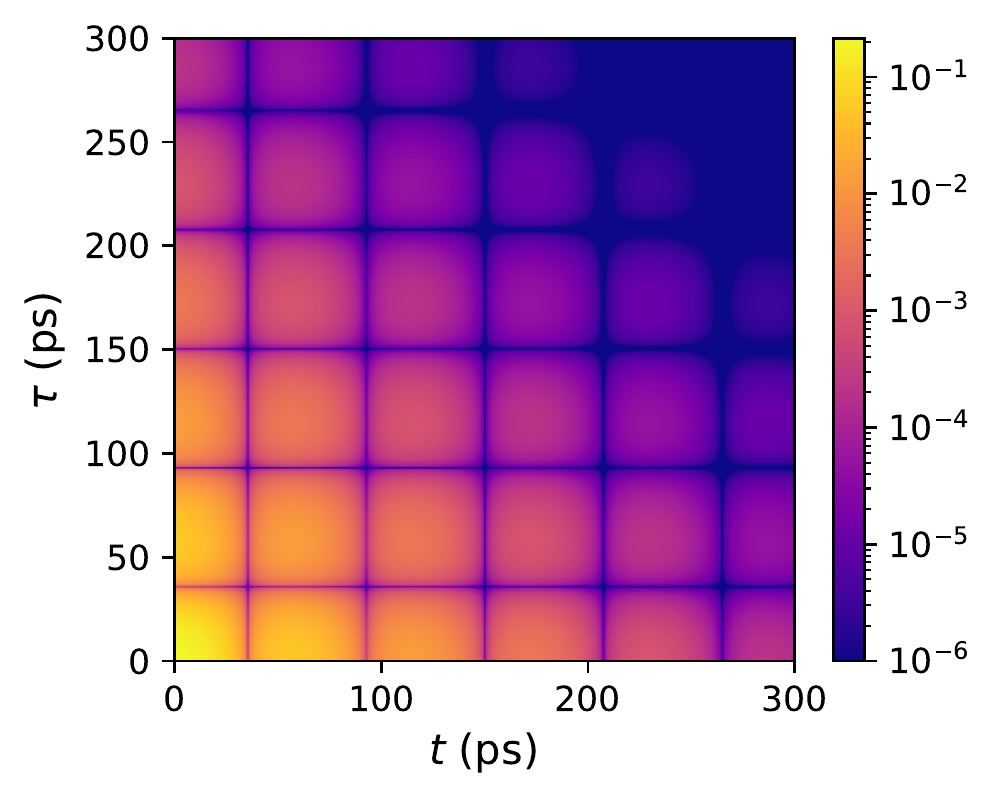}}
\subfloat[]{\includegraphics[width=0.235\textwidth]{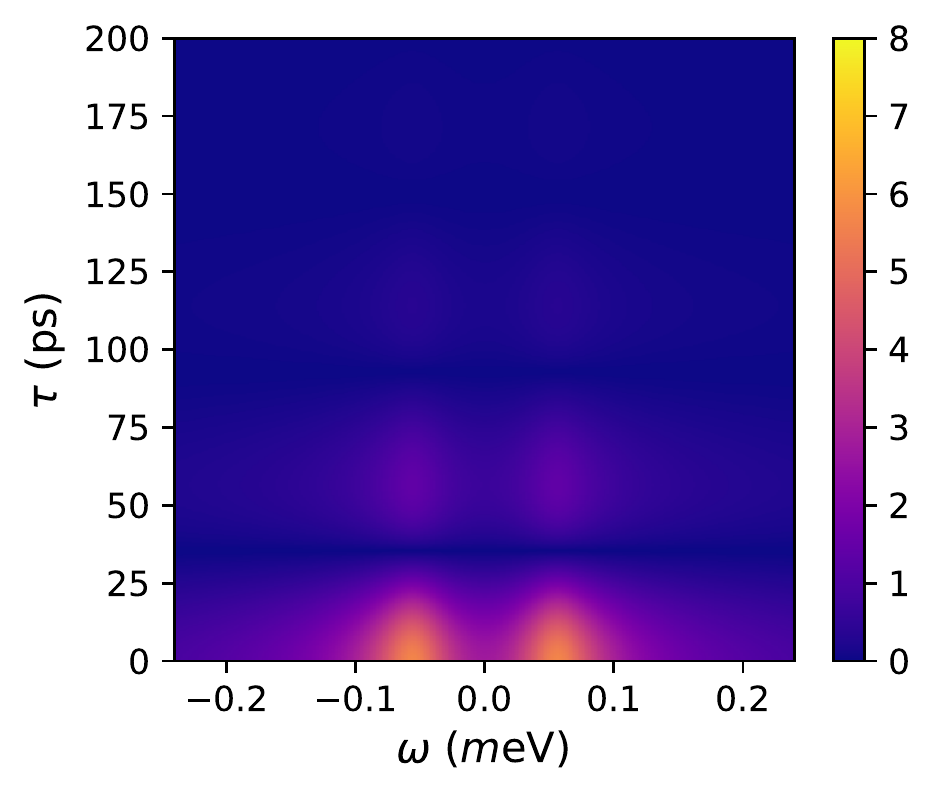}}
\subfloat[]{\includegraphics[width=0.25\textwidth]{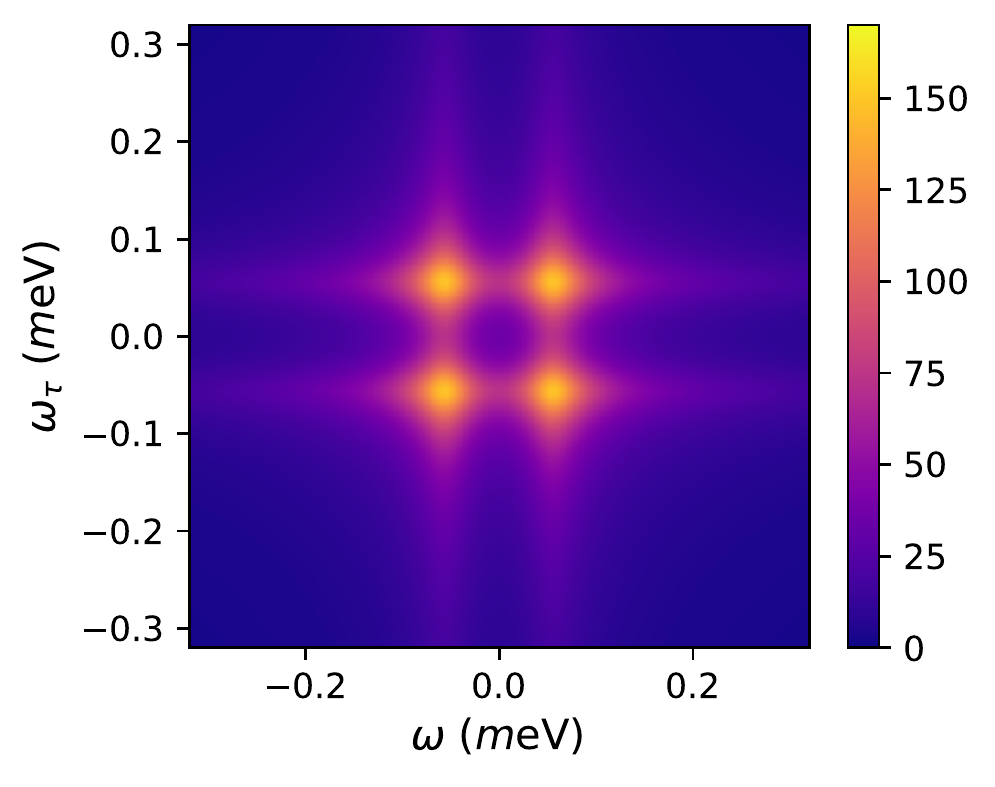}}
\subfloat[]{\includegraphics[width=0.25\textwidth]{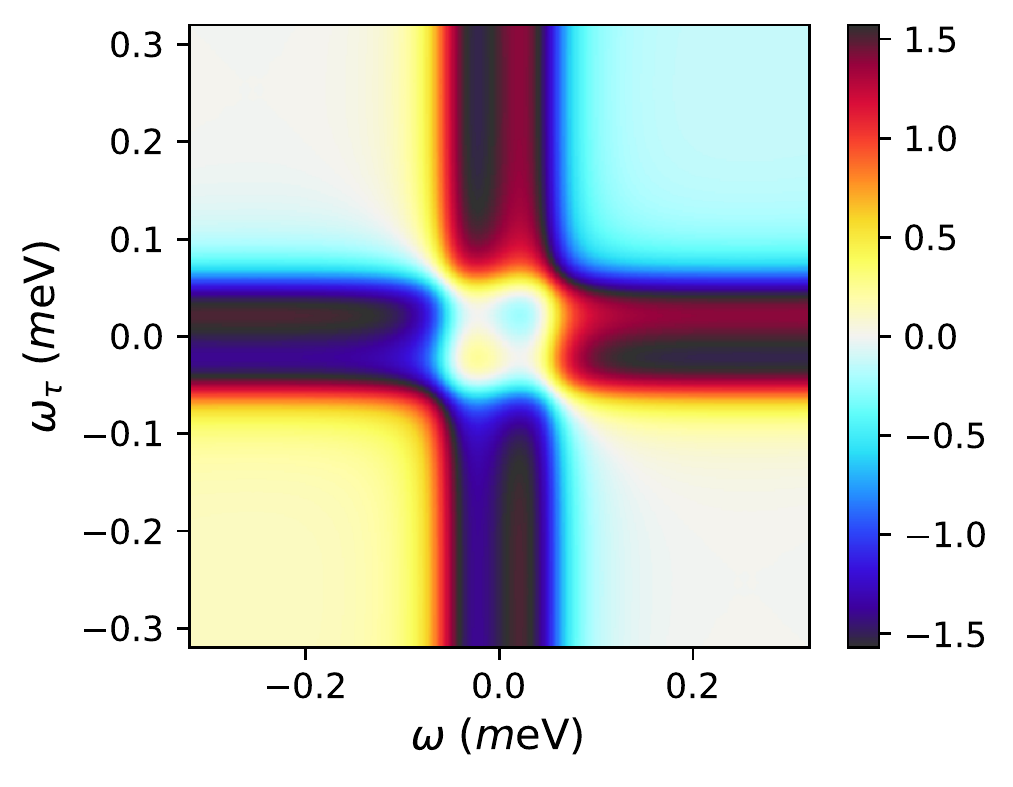}}

\subfloat[]{\includegraphics[width=0.25\textwidth]{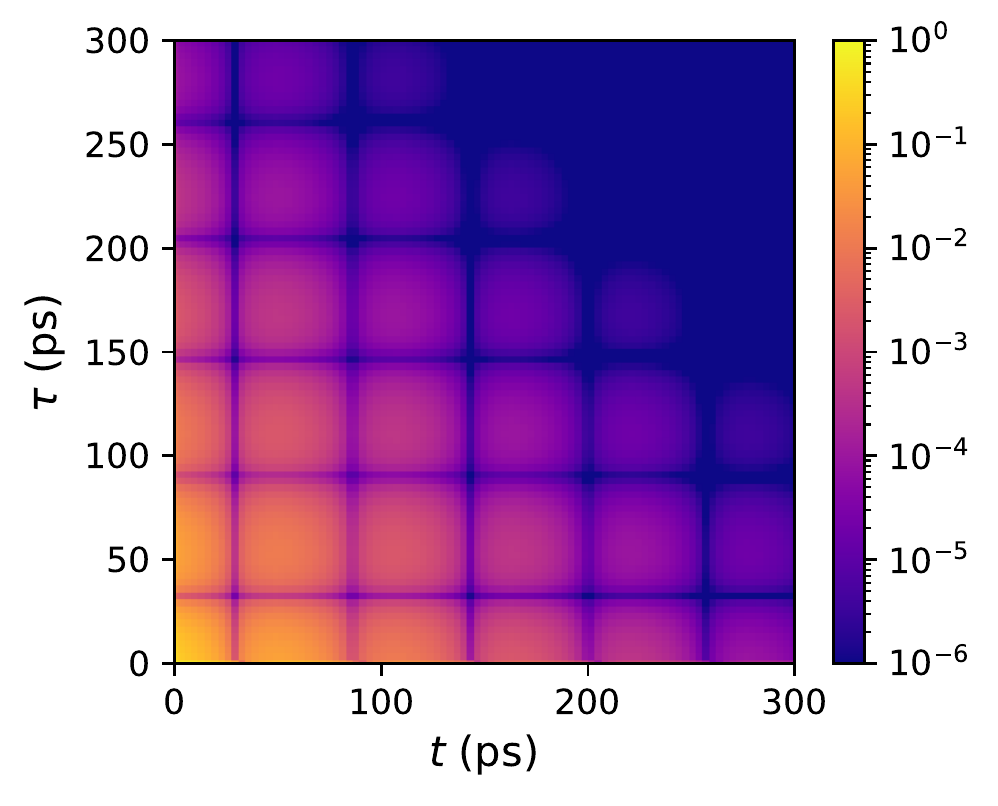}}
\subfloat[]{\includegraphics[width=0.235\textwidth]{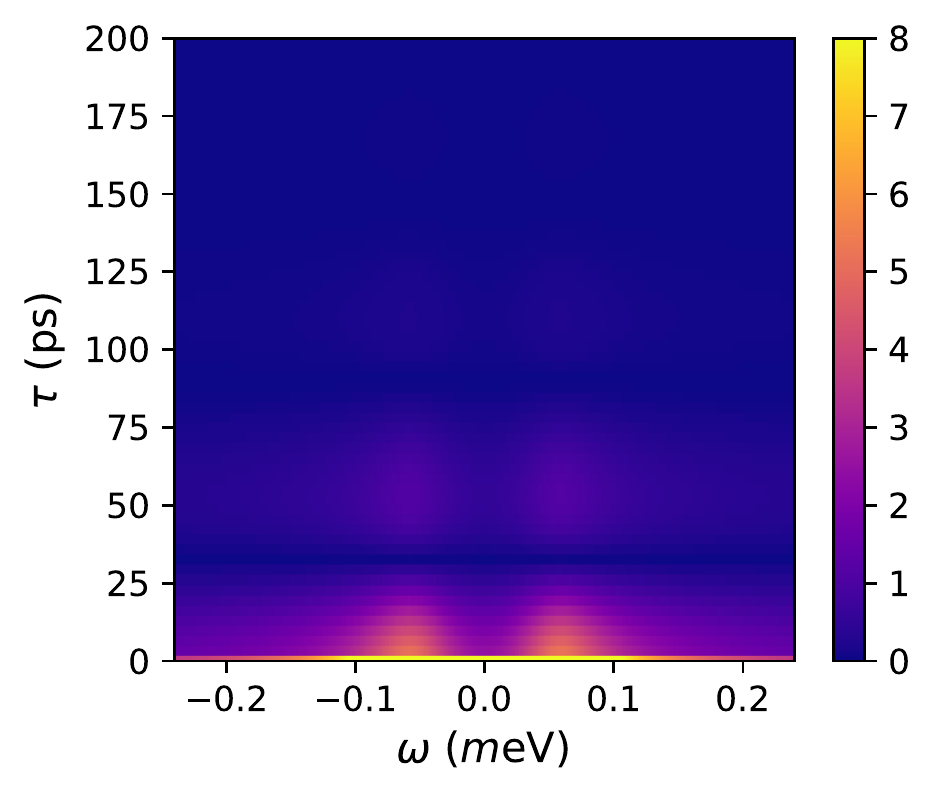}}
\subfloat[]{\includegraphics[width=0.25\textwidth]{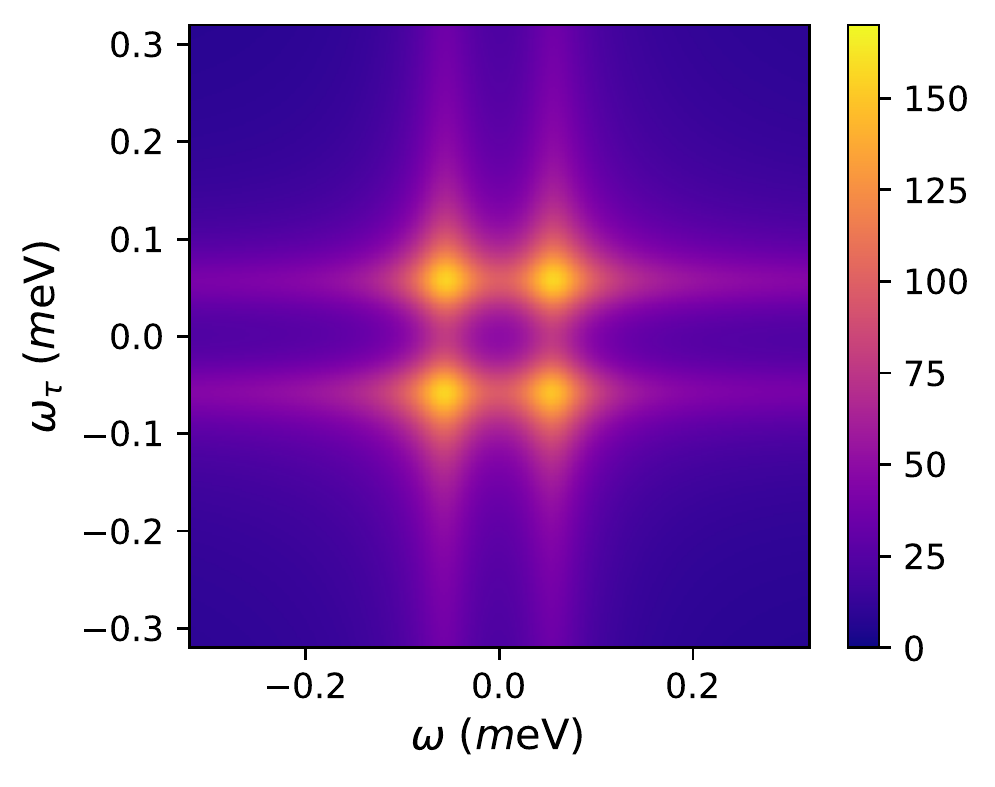}}
\subfloat[]{\includegraphics[width=0.25\textwidth]{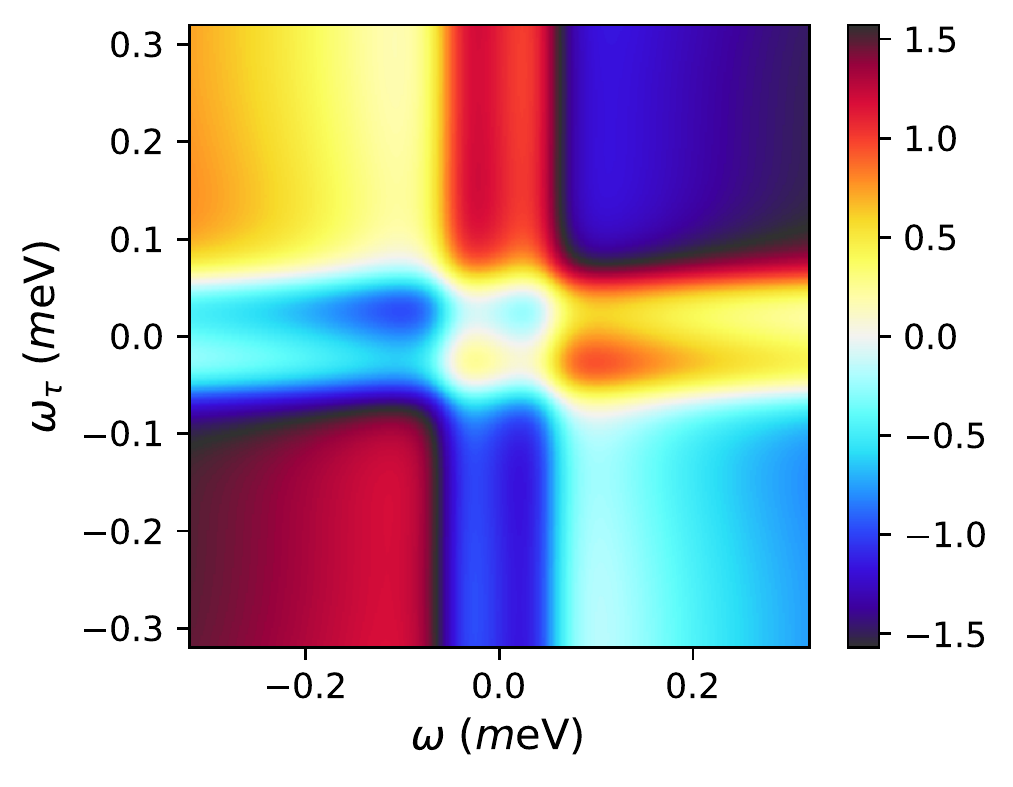}}
\caption{As Fig.\ref{cc-t-tau}, but for the excitation and the measurement both done via the excitonic mode.}
\label{xx-t-tau}
\end{figure}
\end{widetext}

Fig.\ref{xx-t-tau} shows the effect of delay and observation times on the FWM signal when the system is excited via the exciton mode. It is presented for comparison with Fig.\ref{cc-t-tau}, where the excitation is made via the cavity mode. Excitonic excitation only allows ground state to first rang transitions, which manifests itself in a simpler dynamics of the exciton-cavity system. The phonon broad band is however more pronounced due to a much larger initial fraction of the excitonic mode. In Fig.\ref{xx-t-tau} (a,e) the positions of the minima form a square grid, there is no relative shift along the delay time axis. All four peaks are of equal intensity in the 2D spectra in Fig.\ref{xx-t-tau} (c,d).
The NN results show much higher intensity at $\tau=0$ than at $\tau \neq 0$ in Fig.\ref{xx-t-tau}(b).

\bibliography{refferences}

\end{document}